\documentclass[reprint,pra,longbibliography,superscriptaddress,nofootinbib]{revtex4-2}

\usepackage{lipsum}
\usepackage{fancyhdr}
\usepackage{graphicx}
\usepackage[caption=false]{subfig}
\usepackage{braket}
\usepackage{booktabs}
\usepackage{multirow}
\usepackage{bm}
\usepackage{mathtools}
\usepackage{amsmath, esint}
\usepackage{amssymb}
\usepackage{hyperref}
\usepackage{esvect}
\usepackage{float}
\usepackage{placeins}
\usepackage{verbatim}
\usepackage{epstopdf}
\usepackage[normalem]{ulem}
\hypersetup{
    colorlinks=true,
    linkcolor=blue,
    filecolor=magenta,      
    urlcolor=cyan,
    citecolor={blue},
    }

\usepackage[dvipsnames]{xcolor}

\newcommand{\PF}{P}

\allowdisplaybreaks

\makeatletter

\usepackage{comment}
\let\wfs@comment@comment\comment
\let\comment\@undefined

\usepackage[final]{changes}
\@namedef{Changes@AuthorColor}{red}
\colorlet{Changes@Color}{red}

\let\wfs@changes@comment\comment
\let\comment\@undefined

\newcommand\comment{%
    \ifthenelse{\equal{\@currenvir}{comment}}
    {\wfs@comment@comment}
    {\wfs@changes@comment}%
}

\makeatother

\begin{document}

\title{Optimizing the chiral Purcell factor for unidirectional single photon emitters in  topological photonic crystal waveguides using inverse design}

\author{Eric Nussbaum}
\email{eric.nussbaum@queensu.ca}
\affiliation{Centre for Nanophotonics, Department of Physics, Engineering Physics and Astronomy, Queen's University, Kingston, Ontario, Canada, K7L 3N6}

\author{Nir Rotenberg}
\email{nir.rotenberg@queensu.ca}
\affiliation{Centre for Nanophotonics, Department of Physics, Engineering Physics and Astronomy, Queen's University, Kingston, Ontario, Canada, K7L 3N6}

\author{Stephen Hughes}
\email{shughes@queensu.ca}
\affiliation{Centre for Nanophotonics, Department of Physics, Engineering Physics and Astronomy, Queen's University, Kingston, Ontario, Canada, K7L 3N6}

\begin{abstract}
We present an inverse design approach to significantly improve the figures-of-merit for chiral photon elements and quantum emitters in topological photonic crystal slab waveguides. Beginning with a topological waveguide mode with a group index of approximately 10 and a maximum forwards or backwards Purcell factor at a chiral point of less than 0.5, we perform optimizations of the directional Purcell factor. We use a fully three dimensional guided-mode expansion method to efficiently calculate waveguide band dispersion properties and modes, while automatic differentiation is employed to calculate the gradient of objective functions. We present two example improved designs: (i) a topological mode with an accessible group index of approximately 30 and a maximum unidirectional Purcell factor at a chiral point greater then 4.5 representing a nearly 10-fold improvement to the Purcell factor, and (ii)  a slow light mode, well away from the Brillouin zone edge with a group index greater then 350 and a maximum unidirectional Purcell factor at a chiral point greater than 45. 

\end{abstract}

\maketitle

\section{Introduction}

Chiral quantum elements play an increasingly important role in quantum optical experiments and technologies
\cite{Lodahl2017} as they enable non-reciprocal devices 
\cite{sollnerDeterministicPhotonEmitter2015,PhysRevA.90.043802,PhysRevX.5.041036,Scheucher2016, pucherAtomicSpincontrolledNonreciprocal2022}, quantum logic gates
\cite{Shomroni2014,2109.03519}
 and networks
\cite{PhysRevLett.117.240501}. One potential route towards scalable, on-chip chiral elements are photonic crystal  waveguides (PCWs)  integrated with semiconductor quantum dots (QDs)~\cite{mangaraoSingleQuantumdotPurcell2007,lauchtWaveguideCoupledOnChipSinglePhoton2012,RevModPhys.87.347, kurumaTopologicallyProtectedSinglePhotonSources}. Photonic crystal waveguides are known to collect nearly all photons emitted by a quantum dot~\cite{arcariNearUnityCouplingEfficiency2014} while enhancing this emission \cite{Holz2020}, providing indistinguishable photons over long time scales
\cite{Uppu2020} and enabling efficient few photon quantum optical nonlinearities
\cite{PhysRevLett.126.023603}. Chiral interactions arise when a QD with circular polarized transition dipoles is placed in a region where the guided mode field is circularly polarized~\cite{PhysRevA.92.063819, PhysRevLett.128.073602}, or when both dipole and field are elliptically polarized \cite{PhysRevLett.128.073602}, with the direction determined by the matching of the signs of the dipole and field polarizations~\cite{youngPolarizationEngineeringPhotonic2015}.
This has been demonstrated in conventional PCWs~\cite{sollnerDeterministicPhotonEmitter2015} and more recently in a topological PCWs~\cite{mehrabadChiralTopologicalPhotonics2020,Su2021}. 

A significant issue for exploiting PCWs in quantum photonic circuits is that they suffer from disorder-induced backscattering,  which is especially severe in the slow light regime~\cite{ofaolainDependenceExtrinsicLoss2007, hughesExtrinsicOpticalScattering2005, pattersonDisorderInducedCoherentScattering2009, pattersonDisorderinducedIncoherentScattering2009, andreaniLightMatterInteraction2007, kuramochiDisorderinducedScatteringLoss2005, petrovBackscatteringDisorderLimits2009} where emission enhancement occurs. Photon transport through topological PCWs is known to be robust against certain, symmetry-preserving defects such as 60 degree bends~\cite{shalaevRobustTopologicallyProtected2019, heSilicononinsulatorSlabTopological2019, mehrabadChiralTopologicalPhotonics2020},  \added{though we stress that they are not immune to disorder-induced backscattering~\mbox{\cite{rosiekObservationStrongBackscattering2022}}.} 
\deleted{and can  reduce scattering losses in general even while enhancing chiral light-matter interactions~\cite{hauffChiralQuantumOptics}}.
\added{However, they do introduce extra flexibility for potentially reducing backscattering~\mbox{\cite{hauffChiralQuantumOptics}} as well as enhancing chiral light-matter interactions, which is the main focus of this work.}
Despite these \replaced{potential}{known} advantages, the question of how much a topological PC waveguide can enhance chiral interactions remains. Traditionally, PCW designs have been improved using {\it brute force} searches across a few parameters~\cite{zhaoReviewOptimizationMethods2015, liSystematicDesignFlat2008, liLowLossPropagation2012, moriWidebandLowDispersion2005, rigalPropagationLossesPhotonic2017}. This has allowed the slow-light properties of PCWs~\cite{mahmoodianEngineeringChiralLightmatter2017, liSystematicDesignFlat2008} and the quality factors of photonic crystal cavities to be significantly improved. More recently, nanophotonic inverse design has enabled further optimization, leading to robust performance compared to devices designed with conventional methods~\cite{moleskyInverseDesignNanophotonics2018, minkovInverseDesignPhotonic2020, piggottInverseDesignDemonstration2015, christiansenTopologicalInsulatorsTopology2019}. Instead of relying on the designers intuition or brute force searches, inverse design optimizes a large number of parameters describing the device, allowing novel and intuitive designs to be obtained~\cite{mannReducingDisorderinducedLosses2013}.

In this work, we show how to optimize a state of the art topological PCW~\cite{mehrabadChiralTopologicalPhotonics2020} to enhance quantum chiral interactions. We use shape optimization~\cite{minkovInverseDesignPhotonic2020}  to demonstrate an order of magnitude improvement to the directional Purcell factor. To achieve this task, we use the efficient guided-mode expansion method (GME)~\cite{andreaniPhotoniccrystalSlabsTriangular2006} to calculate the band structure and modes of PCWs, and automatic differentiation to calculate the gradient of objective functions. In Sec.~\ref{sec:Theory}, we outline the GME method, discuss how the directional Purcell factor is calculated, and how we apply inverse design. In Sec.~\ref{sec:Results}, the results of two optimizations are presented, one operating at a moderate group index $n_g \approx 30$ and the other operating at a larger group index $n_g \approx 350$. In both cases, single mode operation, well away from the Brillouin zone edge, is maintained. Finally, we conclude in Sec.~\ref{sec:Conclusions}. 

\section{General Theory and Optimization Protocol}
\label{sec:Theory} 

In this section, we describe how we can \added{use GME to} efficiently and rapidly optimize complex, topological PCWs without the need for computationally intensive methods such as finite element or finite-difference 
time-domain (FDTD) techniques \added{which are are at least one order of magnitude slower then the GME method for calculating band structures and Bloch modes. For example, full photonic band structure calculations using GME can be performed in a few minutes on a standard desktop computer, which can take around one day using COMSOL.} 

\replaced{After discussing the GME method for obtaining PCW modes and the band structure, we then}{
We briefly describe how the GME method allows us to obtain PCW modes and band structure efficiently, then} show how the directional Purcell factor is calculated, and finally discuss how we apply inverse design to rapidly improve chiral figures of merit.

\added{
For high-index-contrast
like (planar) PCWs, the
GME is highly accurate for computing the dispersion. 
 Moreover, the GME can also be used
to obtain the radiative losses for modes above the light line,
using perturbation theory 
\mbox{\cite{andreaniLightMatterInteraction2007,andreaniPhotoniccrystalSlabsTriangular2006,sauerTheoryComputationIntrinsic2021,sauerTheoryComputationIntrinsic2021}}.
In the structures we use below, we only consider the practical area of
below the light line modes which formally have no intrinsic losses, in the absence of disorder. The role of disorder will be investigated in future work.
It is also worth mentioning that
the GME can also be used to obtain accurate cavity mode losses~\mbox{\cite{andreaniPhotoniccrystalSlabsTriangular2006}},
and can be used as input to a Bloch mode expansion for computing full 3D Anderson localization modes~\mbox{\cite{vascoStatisticsAndersonlocalizedModes2017}}.}

\subsection{Guided-mode expansion method}
The GME solves Maxwell's equations by expanding the magnetic field in the basis of the modes of the effective homogeneous slab, and then solves the resulting eigenvalue problem~\cite{andreaniPhotoniccrystalSlabsTriangular2006}.
The GME is a 3D vectorial method and is significantly faster than alternative methods such as
FDTD for calculating the band structure and modes of PCWs~\cite{sauerTheoryIntrinsicPropagation2020}. In this work,  we use a GME Python library Legume of Minkov \textit{et al.}~\cite{minkovInverseDesignPhotonic2020, FancomputeLegume2020}. The main details of the GME method are summarized below. 

Within a linear dielectric medium, Maxwell's equations for the electric and magnetic fields, $\bm E(\bm r, t)$ and $\bm H (\bm r, t)$ respectively, can be rewritten in the frequency domain to obtain an eigenvalue equation.
In terms of the magnetic field, $\bm H(\bm r, \omega)$, one has:
\begin{equation}
    \label{eq: maxwell}
    \bm \nabla \times \left( \frac{1}{\epsilon(\bm r)} \bm \nabla \times \bm H(\bm r, \omega) \right)  = \left(\frac{\omega}{c}\right)^2 \bm H(\bm r, \omega),
\end{equation}
where $\epsilon(\bm r)$ is the dielectric constant, with the condition $\nabla \cdot \bm H(\bm r, \omega) =0$. 
We assume lossless media
and neglect frequency dispersion in the dielectric constant, which is a good assumption for the materials of interest
for designing and making PCW waveguide modes.
Once the magnetic field is calculated, the electric field is easily obtained from
\begin{equation}
    \label{eq: E from H}
    \bm E(\bm r, \omega) = \frac{i}{\omega \epsilon_0 \epsilon(\bm r)} {\bm \nabla} \times \bm H (\bm r, \omega).
\end{equation}

To solve equation~\eqref{eq: maxwell},  the GME method expands the magnetic field into an orthonormal set of basis states as
\begin{equation}
    \label{eq:orthonormal basis states}
    \bm H(\bm r, \omega) = \sum_\mu c_\mu \bm H_\mu (\bm r),
\end{equation}
and so equation~\eqref{eq: maxwell} can be written as
\begin{equation}
    \label{eq:eigen}
    \sum_\nu \mathcal{H}_{\mu \nu}c_\nu = \frac{\omega^2}{c^2} c_\mu,
\end{equation}
where the elements of the Hermitian matrix $\mathcal{H}_{\mu \nu}$ are defined as
\begin{equation}
    \label{eq:eigenMatrix}
    \mathcal{H}_{\mu \nu} = \int \cfrac{1}{\epsilon(\bm r)} \left({\bm \nabla}  \times \bm H^{*}_\mu(\bm r) \right) \cdot \left(  {\bm \nabla}  \times \bm H_\nu(\bm r) \right) d \bm r.
\end{equation}

To define an appropriate basis set $\bm H_\mu (\bm r)$, the GME method uses the guided modes of the effective homogeneous slab waveguide, with a dielectric constant taken as the spatial average of the dielectric constant in the slab layer of the PCW being studied~\cite{andreaniPhotoniccrystalSlabsTriangular2006}.
The guided modes of the homogeneous slab depend on a wave vector, which can take any value in the slab plane, while the modes of the PCW depend on $\bm k$, which we restrict to the first Brillouin zone. Thus, for the magnetic field at each wave vector, only the effective waveguide modes with wave vector $\bm k + \bm G$, are included in the basis. The guided mode expansion for the Bloch modes is then
\begin{equation}
    \label{eq:GME magnetic field}
    %\bm H_{\bm k} (\bm r) = \sum_{\bm G, \alpha} c(\bm k + \bm G, \alpha) \bm H_{\bm k + \bm G , \alpha} ^{\rm guided} (\bm r),
    \bm H_{\bm k} (\bm r) = \sum_{\bm G, \alpha} C_{\bm k + \bm G, \alpha} \bm H_{\bm k + \bm G , \alpha} ^{\rm guided} (\bm r),
\end{equation}
where $\bm H_{\bm k + \bm G , \alpha} ^{\rm guided} (\bm r)$ is a guided mode (which is known analytically) of the effective waveguide and $\alpha$ is the index of the guided mode. Since the dielectric is periodic both the magnetic and electric fields can be written according to Bloch's theorem.

The electric and magnetic field Bloch mode can be written as
\begin{align}
    \bm E_{\bm k} (\bm r) &= \bm e_{\bm k}(\bm r) e^{i \bm k \cdot \bm r},\\
    \bm H_{\bm k} (\bm r) &= \bm u_{\bm k}(\bm r) e^{i \bm k \cdot \bm r},
\end{align}
where $\bm e_{\bm k}(\bm r)$ and $\bm u_{\bm k}(\bm r)$ are the electric and magnetic field Bloch modes for wave vector $\bm k$, which are normalized according to $\int_{\rm cell} \epsilon(\bm r) |\bm e_{\bm k}(\bm r)|^2 d\bm r = 1$ throughout this paper, and the integration is carried out over one unit cell.
Note, the GME can also be used to compute intrinsic losses for Bloch modes above the light line~\cite{andreaniGuidedModeExpansionMethod2006,sauerTheoryComputationIntrinsic2021}.

\subsection{Purcell factor for a single quantum emitter}
The coupling between a point-dipole quantum emitter and a waveguide mode can be calculated from a Green's function analysis~\cite{mangaraoSingleQuantumdotPurcell2007, youngPolarizationEngineeringPhotonic2015}. The Green's function for the waveguide mode, describing the response at $\bm r$ to an oscillating dipole at $\bm r_0$, is
\begin{equation}
\label{eq: Green's fxn}
\begin{aligned}
    \bm G_{w}(\bm r, \bm r_0,\omega) &= \bm G_{f}(\bm r_0, \bm r,\omega) + \bm G_{b}(\bm r, \bm r_0,\omega)\\
    &= \frac{i a \omega}{2 v_g} [\Theta (x- x_0) \bm e_k(\bm r) \otimes \bm e^*_k(\bm r_0) e^{ik(x-x_0)} \\ 
    &+ \Theta (x_0 - x) \bm e_k^*(\bm r) \otimes \bm e_k(\bm r_0) e^{-ik(x-x_0)}],
\end{aligned}
\end{equation}
where $v_g$ is the group velocity and $\Theta$ is the Heaviside step function.  The subscripts
refer to backward (`$b$')
and forward (`$f$') mode contributions.

The rate of photon emission, from a QD exciton into the waveguide mode can be split into forwards and backwards rates. Representing the dipole as a unit vector $\hat{\bm \mu}$ with dipole moment $d_0$, the forwards and backwards emission rates are~\cite{youngPolarizationEngineeringPhotonic2015}, respectively:
\begin{align}
    \Gamma_w^f (\bm r_0) &= {\rm Im} \left[\frac{2 d_0^2 \hat{\bm \mu}^\dagger \cdot \bm G_{f}(\bm r_0, \bm r_0,\omega) \cdot \hat{\bm \mu}}{\hbar \epsilon_0}\right], \\
    \Gamma_w^b (\bm r_0) &= {\rm Im} \left[ \frac{2 d_0^2 \hat{\bm \mu}^\dagger \cdot \bm G_{b}(\bm r_0, \bm r_0,\omega) \cdot \hat{\bm \mu}}{\hbar \epsilon_0}\right].
    \label{eq: Gamma f and b}
\end{align}
These expressions are valid for lossless media \cite{PhysRevLett.127.013602}, otherwise
a more careful quantum analysis is needed, where the rate
of emission is no longer proportional to the projected 
local density of states, and one must 
include additional contributions from vacuum fluctuations.

The Purcell factor is defined as $\PF = \Gamma_w/\Gamma_0$, where 
\begin{equation}
    \Gamma_0 = \frac{\omega^3 d_0^2 \sqrt{\epsilon_s}}{3 \hbar \pi \epsilon_0 c^3},
\end{equation}
is the emission rate into the corresponding homogeneous medium with dielectric constant $\epsilon_s$ \cite{mangaraoSingleQuantumdotPurcell2007}. Separating the Purcell factor into its forwards and backwards components, 
\begin{equation}
\label{eq: PF f and b}
\begin{aligned}
   \PF^f (\bm r_0) &= \frac{3\pi c^3 a}{v_g \omega^2 \sqrt{\epsilon_s}}\hat{\bm \mu}^\dagger \cdot \left [\bm e_k(\bm r_0) \otimes \bm e_k^*(\bm r_0)\right ] \cdot \hat{\bm \mu} \\
    \PF^b (\bm r_0) &= \frac{3\pi c^3 a}{v_g \omega^2 \sqrt{\epsilon_s}}\hat{\bm \mu}^\dagger \cdot \left [\bm e_k^*(\bm r_0) \otimes \bm e_k(\bm r_0) \right] \cdot \hat{\bm \mu},
\end{aligned}
\end{equation}
where $v_g$ and ${\bm e}_k$ implicitly depends on frequency.
%and, e.g., becomes much smaller closer to
%a mode edge.

At a point where the field of the forward propagating mode is right circularly polarized (RCP),
with unit vector 
$\hat{\bm n}_R = 2^{-1/2}(\hat{\bm x}+i \hat{\bm y})$,
a RCP dipole $\bm \mu = \bm \sigma_+$ will only couple to this mode and not to the one propagating backwards; similarly,  a left circularly polarized (LCP) dipole $\bm \mu = \bm \sigma_-$ placed at the same position, will only couple to the backwards propagating mode~\cite{youngPolarizationEngineeringPhotonic2015}. 

With a focus on optimizing chiral photon elements, we also define the {\it chiral directionality},
\begin{equation}
    C = \frac{\PF^f - \PF^b}{\PF^f + \PF^b},
\end{equation}
as well as the {\it chiral forwards Purcell factor},
\begin{equation}
    \tilde{\PF}^f = C \PF^f,
    %\widetilde{\PF}^f = C \PF^f,
    %\PF^f_C = C \PF^f,
\end{equation}
which will be used below as representative figures of merit.

\subsection{Inverse design methodology}

Inverse design treats the design process as an optimization problem. Instead of using a heuristic understanding of a specific phenomenon and adjusting a small number of related parameters, inverse design optimizes a large number of device parameters. This can lead to unintuitive designs with significantly better performance then their traditionally designed counter parts~\cite{moleskyInverseDesignNanophotonics2018}. 

Gradient based optimization algorithms are usually used in inverse design. To efficiently calculate objective function gradients the adjoint variable method is widely used~\cite{minkovInverseDesignPhotonic2020}. However, implementing the adjoint variable method can be extremely difficult with a solver which solves an eigenvalue problem, such as the GME method~\cite{minkovInverseDesignPhotonic2020}. Instead, automatic differentiation can be used to calculate the gradient of an arbitrarily complex objective function. An automatic differentiation library tracks all sub-function executions, for which the gradient is known, and then uses the chain rule to compose the objective functions gradient~\cite{minkovInverseDesignPhotonic2020}. In this work, we use the automatic differentiation Python library Autograd~\cite{maclaurinHIPSAutograd}. The GME library we use, `Legume', was developed to be compatible with Autograd making it straightforward to calculate the gradient of the objective functions based on a GME calculation~\cite{minkovInverseDesignPhotonic2020}.

Starting from an initial design, shape optimization modifies the boundaries between different materials in order to optimize the objective function. In comparison, topology optimization places fewer constraints on how the material distribution can evolve, including allowing the number of holes to be changed. Shape optimization is therefore well suited to optimizing designs where the general shape is known before hand~\cite{michaelsLeveragingContinuousMaterial2018} while topology optimization can produce remarkably inventive designs.

\section{Results}
\label{sec:Results} 

Beginning with a state-of-the-art topological PCW~\cite{mehrabadChiralTopologicalPhotonics2020}, we will show several improved designs. The optimizations are performed in two phases: first the maximum $\tilde{\PF}^f$ is optimized for a particular mode and wave vector, and second,  the dispersion is optimized for a slow light region. One design is optimized to operate at an accessible group index $n_g = c/v_g$ of approximately 30, while the other has a region of very slow light with a group index of greater then 350 away from the Brillouin zone edge.

\subsection{Initial design}

\begin{figure}
    \centering
    \includegraphics[width=\linewidth]{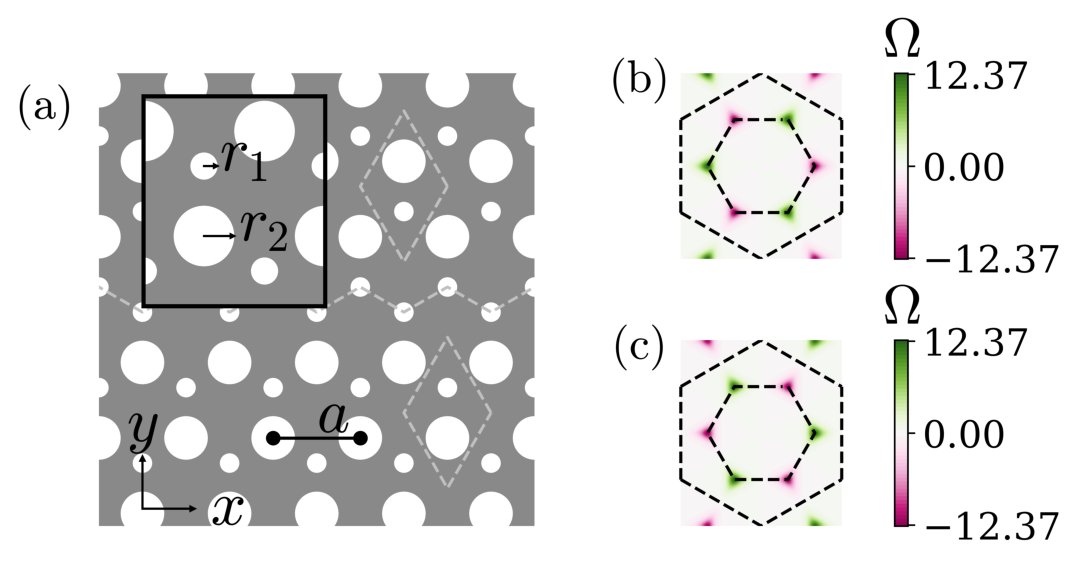}
    \caption{(a) Schematic of the topological edge-mode waveguide. The direction of propagation $x$ and relevant geometric parameters are indicated. (a) Schematic (top-down view) of the crystal. The rhombic unit cell is outlined. Magnitude of the Berry curvature for the mode bellow the band gap ${ \Omega}(\bm k)$ (units of $[a^2/4 \pi^2]$) 
    %\en{i.e. the Berry curvature with units $[a^2 / 4 \pi]$ not $\Omega \times a^2/4\pi^2$, is this clear in the uncolored text?} % seems fine to me- SH
    for the upper (b) and the lower (c) crystals shown in (a) for a slab thickness $h = 170$~nm, $a=250$~nm, $r_1 = 28$~nm, $r_2=62.5$~nm. The outer dashed black line is the edge of the first Brillouin zone and the inner dashed line's vertices are the $K$ and $K'$ points.
    }
    \label{fig: berry}
\end{figure}

The optimization is started from a reference topological PCW in which chiral coupling of single QDs to the waveguide mode has been experimentally realized~\cite{mehrabadChiralTopologicalPhotonics2020}, using a promising design. The PCW modes are formed at the interface between two topologically distinct crystals. The two crystals are inversion symmetry partners and are formed from a hexagonal lattice of unit cells containing two circular holes of different sizes, with radii $r_1$ and $r_2$, as shown in Fig.\ref{fig: berry}a. \added{Similar designs using triangular holes ~\mbox{\cite{yoshimiSlowLightWaveguides2020}} have also been fabricated~~\mbox{\cite{yoshimiExperimentalDemonstrationTopological2021, kurumaTopologicallyProtectedSinglePhotonSources, barikTopologicalQuantumOptics2018}} and our approach could be applied to these designs as well. However,  we have focused on a design with circular holes as these are typically easier to fabricate with small dimensions.}
 
In analogy with topological electronics, the topology of these crystals is related to their Chern number~\cite{luTopologicalPhotonics2014} which due to the time-reversal symmetry of this photonic system, is equal to zero. However, this need not be the case for the valley Chern index which we calculate by partitioning the integration of the Berry curvature 
\begin{equation}
 \bm \Omega_n(\bm k) = \bm \nabla_{\bm k} \times i \Braket{ \bm u_{\bm k, n} | \bm \nabla_{\bm k} | \bm u_{\bm k, n}},
\end{equation}
where $n$ is the index of the mode,
over the Brillouin zone to around only the $K$ or $K'$ valleys~\cite{heSilicononinsulatorSlabTopological2019}, as shown in figure~\ref{fig: berry}b,c. For details on how to numerically calculate the Berry curvature see Ref.~\cite{pazTutorialComputingTopological2020}. The valley Chern index is $C_V = C_K - C_{K'}$ where, for example, $C_K = \frac{1}{2\pi} \oiint_{ BZ,K} \Omega_n(\bm k) dk^2$ is the integral of the Berry curvature over half the Brillouin zone around the $K$ valleys. Unlike the Chern number, the valley Chern index is not necessarily quantized~\cite{heSilicononinsulatorSlabTopological2019, pazTutorialComputingTopological2020}. The two crystals on each side of the interface have the same $|C_V|$ though with opposite sign. 
This opposing sign gives rise to valley Hall edge states which are expected, and have been shown to be more robust against defects that do not couple the counter-propagating states~\cite{shalaevRobustTopologicallyProtected2019, heSilicononinsulatorSlabTopological2019}.

\begin{figure}[htb!]
    \centering
    \includegraphics[width=\linewidth]{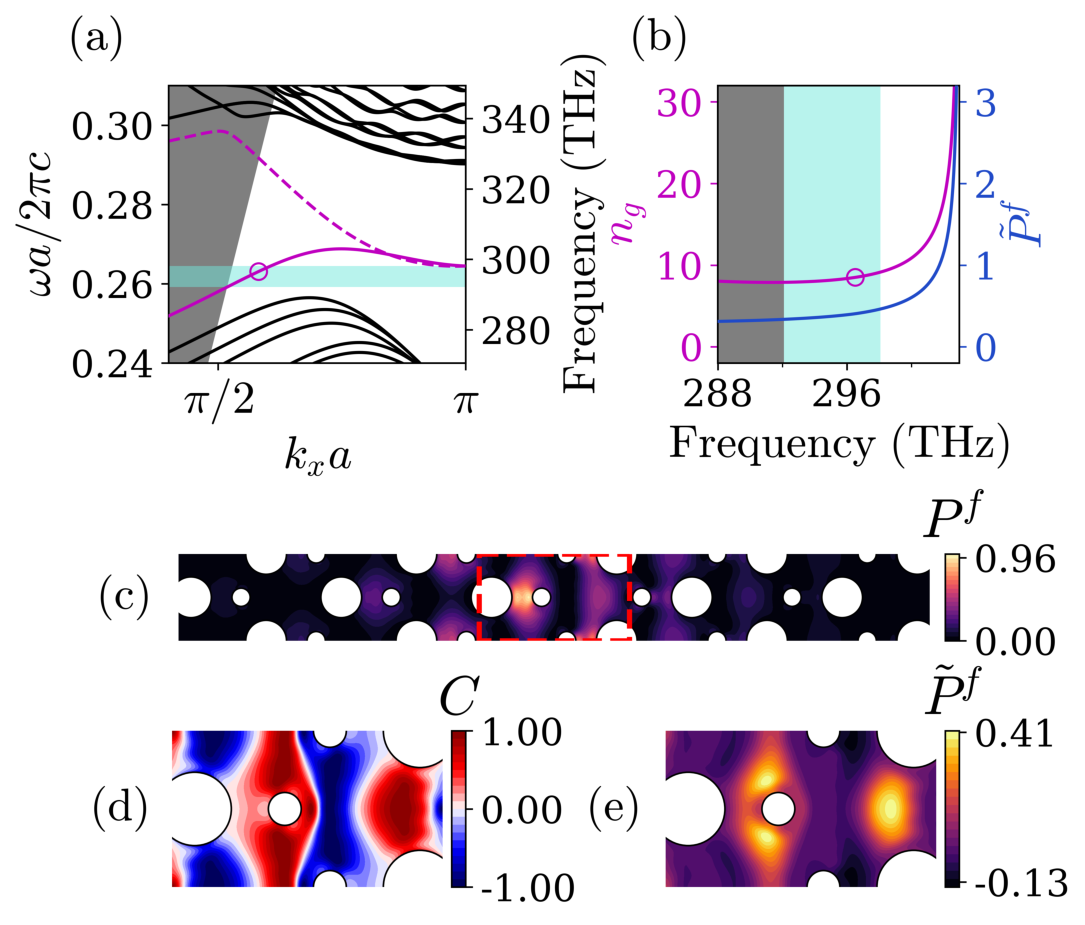}
    \caption{
    (a) Band structure for the initial design~\cite{mehrabadChiralTopologicalPhotonics2020} with $a = 266$~nm. The (NT) guided mode is drawn in solid purple, the second guided mode in dashed purple, and bulk modes in black. The NT mode's SM bandwidth is highlighted in turquoise. Additional guided modes existing at the secondary interface at the edge of the GME unit cell have been removed in this figure, as well as in figures~\ref{fig: NDBP bands} and \ref{fig: allSMng bands} (see Appendix~\ref{sec: AppA}). (b) The group index for the NT mode is drawn in purple and the maximum chiral forwards Purcell factor is drawn with a solid blue line. Frequencies at which the NT mode is above the light line are highlighted in grey. The following fields are calculated at the circular marker in (a) and (b) where $n_g = 8.5$. The $xy$-plane cross sections are at the center of the slab. (c) Forwards Purcell factor $\PF^f$ for $\bm \mu = \bm \sigma_-$, (d) the associated chiral directionality $C$, and (e) the associated chiral forwards Purcell factor $\tilde{\PF}^f$. The red dashed box in (c) encloses the region shown in (d) and (e).
    }
    \label{fig: seed bands}
\end{figure}

This initial topological waveguide is designed to operate near a wavelength of 1000 nm in a 170-nm thick GaAs (dielectric constant of 11.6) membrane, and so has a pitch $a=266$~nm, $r_1 = 28$~nm, and $r_2=62.5$~nm~\cite{mehrabadChiralTopologicalPhotonics2020}. As shown in figure~\ref{fig: seed bands}a, the initial design has two guided modes in the frequency range of interest. We refer to these modes as trivial (T) and non-trivial (NT), and in what follows focus on the latter. This is because the NT mode has been experimentally demonstrated to support chiral coupling to single QDs and used to form triangular ring resonators~\cite{mehrabadChiralTopologicalPhotonics2020}. Further, as demonstrated in Ref.~\cite{mehrabadChiralTopologicalPhotonics2020}, when a corner is placed in the waveguide the NT mode propagates around the corner while the T guided mode is prone to backscattering. We note that, if we begin with the bearded interface in figure~\ref{fig: berry}(a) and increase the radius of one of the rows of small holes at the interface to the size of the large holes such that there is a smooth transition to a zigzag interface, then the NT mode remains throughout, while the other guided mode disappears into the bulk bands~\cite{mehrabadChiralTopologicalPhotonics2020}.

For this initial design, we observe that the NT edge state has a single mode (SM) bandwidth positioned around 296~THz. Within this frequency band, at 296.5~THz, at a spectral separation of 1.6 THz away from the band edge, the mode has a group index $n_g = 8.5$, as shown in Fig.~\ref{fig: seed bands}b by the circular marker. Although chiral emission into this structure has been observed~\cite{mehrabadChiralTopologicalPhotonics2020}, the calculated $\PF^f$ map peaks at 0.96 (Fig.~\ref{fig: seed bands}c), demonstrating that this emission is in fact suppressed. Worse, the maxima of $\PF^f$ and $C$ do not overlap for this initial design, and we find a maximal $\tilde{\PF}^f = 0.41$ (Fig.~\ref{fig: seed bands}e). In the region where $C \geq 0.97$, the maximum forwards Purcell factor is $\max(\PF^f) = 0.41$.

Before optimizing the design, we shift the pitch to  $a=250$~nm to increase the operational frequency to ensure that the optimized design can function around $320$~THz, the desired frequency for working with high-quality gallium arsenide quantum dots \cite{lodahlInterfacingSinglePhotons2015}. 
However, because of fabrication limits, we do not decrease $r_1$ bellow our minimum allowed size of $r_{\min} = 27.5$~nm~\cite{mehrabadChiralTopologicalPhotonics2020} so we use the same hole sizes in $r_1 = 28$~nm and $r_2=62.5$~nm with $a=250$~nm. We note that changing the hole radii by this amount  does not adversely affect the concentration of the Berry curvature with opposite signs at the $K$ and $K'$ points. After the optimization, the periodicity can be increased if it is necessary to decrease the operational frequency, but the periodicity can not in general be further decreased without violating our fabrication-imposed constraints.

\subsection{Optimization}
One of our objectives for the optimization is to improve the system's performance as a bright, or rapid, chiral light-matter interface. We therefore demand single mode operation, and only consider positions that are at least 35~nm away from the nearest hole edge, to ensure that the QDs can be electrically gated~\cite{pregnolatoDeterministicPositioningNanophotonic2020}. The optimized design should have an accessible group index of less than 50 with a Purcell factor, in a region where $C \approx 1$, of approximately 5 which is close to the optimum Purcell factors for enabling efficient pulse-triggered single photon sources~\cite{PhysRevB.98.045309,Somaschi2016,PhysRevLett.116.020401}. 
As discussed above and shown in figure~\ref{fig: seed bands}, the maximum chiral  Purcell factor for the initial design is 0.41. We are therefore targeting an increase by a factor of approximately 10. With the same restrictions, apart from allowing for a large group index greater then 50, we also seek a design with a maximized chiral forwards Purcell factor.

A shape parameterization is used where the position and radii of the 12 holes closest to the interface are allowed to be modified. To ensure the final design can be fabricated extremely small features must be avoided. Therefore the minimum allowed hole radius is 27.5~nm and the minimum allowed distance between hole edges is 40~nm. At the start of each optimization, iteration a projection is applied to the design parameters which transforms designs which violate theses constraints to a similar design which does not. The ``Adam''~\cite{kingmaAdamMethodStochastic2017} optimization algorithm is used throughout.

A two-step optimization is performed. The first optimization phase is focused on improving $\max(\tilde{\PF}^f)$ in a region approximately 35~nm away from the nearest hole edge. The objective function for the first phase, is
\begin{equation}
    \label{eq:F1}
    \mathcal{F}_1(\bm \gamma) = {\rm max}[G(\bm r) \tilde{\PF}^f (\bm r)] G_{\rm SM},
\end{equation}
where $\bm \gamma$ is a vector of the device parameters, $G(\bm r)$ is a smooth step function that is equal to 0 inside a hole and approximately 1 more then 35~nm from a hole,
\begin{equation}
    G(\bm r) = \dfrac{1}{1 + e^{-g(r_m(\bm r) - r_s)}},
\end{equation}
where $r_m(\bm r)$ is the minimum distance to a hole edge, $r_s = 30$~nm and $g = 74~{\rm nm}^{-1}$ are parameters, $G \tilde{\PF}^f$ are calculated for the NT mode at a fixed wave vector, and $G_{\rm SM}$ encourages a certain magnitude and position of SM bandwidth.
The objective function for the first phase is maximized.

The intermediate design from the first phase is further optimized in a second phase of optimization which is focused on improving the dispersion characteristics for slow-light operation. Two different objective functions are used to obtain two different styles of design. The first is designed to obtain our primary objective, of operation at an accessible group index,
\begin{equation}
    \label{eq:F2A}
    \mathcal{F}_{2A}(\bm \gamma) = \int_{k_{x,0}}^{k_{x,1}} d k_x (n_g - n_{g,d})^2, 
\end{equation}
where $n_g$ is the group index of the NT mode, $n_{g,d} = 30$ is the desired group index, and the integral is performed over $k_x a = [0.6, 0.9]\pi$.
The second, is used to obtain a very large $n_g$ inside the NT mode's SM bandwidth,
\begin{equation}
    \label{eq:F2B}
    \mathcal{F}_{2B}(\bm \gamma) = \omega_{b} - \omega_{{\rm SM}, \max} + \frac{D}{\max_{\rm SM}(n_g)}   
\end{equation}
where $\omega_{b}$ is the frequency, under the light line, where $n_g$ first changes sign, $\omega_{{\rm SM}, \max}$ is the maximum frequency of the NT mode's lowest frequency region of SM bandwidth, $D$ is a weight, and $\max_{\rm SM}(n_g)$ is the maximum $n_g$ in the NT mode's lowest region of SM bandwidth. OF B therefore works to make the NT mode's bandwidth entirely SM and to increase the maximum $n_g$ in the SM bandwidth. Both $\mathcal{F}_{2A}(\bm \gamma)$ and $\mathcal{F}_{2B}(\bm \gamma)$ are minimized.

\subsection{Improved designs}
\begin{figure}[htb!]
    \centering
    \includegraphics[width=\linewidth]{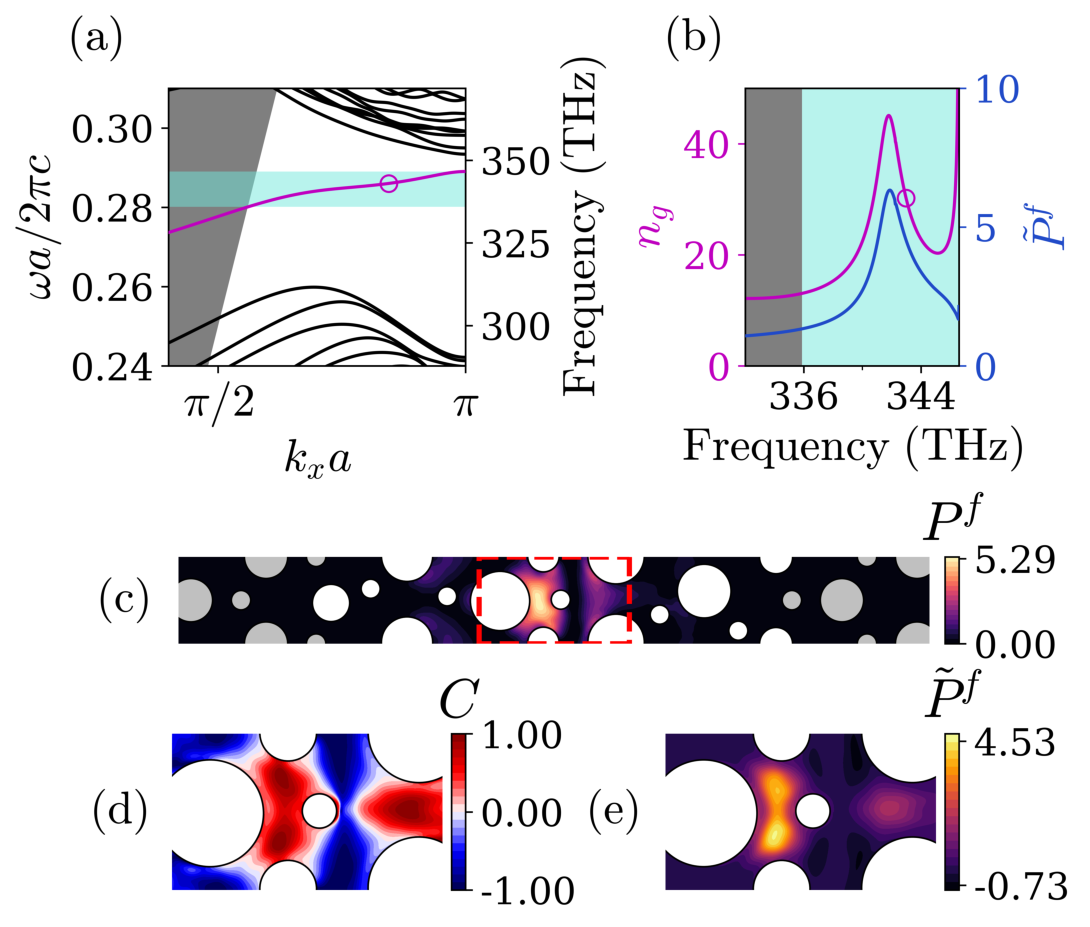}
    \caption{
    Summary of new design A.
    (a) Band structure. (b) Group index and the maximum chiral forwards Purcell factor for the NT mode. The following fields are calculated at the circular marker in (a) and (b) where $n_g = 30.2$. (c) Forwards Purcell factor $\PF^f$ for $\bm \mu = \bm \sigma_-$, (d) the associated chiral directionality $C$, and (e) the associated chiral forwards Purcell factor $\tilde{\PF^f}$. The holes which were not allowed to be modified during the optimization are drawn in grey while the holes which were are drawn in white.
    }
    \label{fig: NDBP bands}
\end{figure}

The results of our first optimization routine, designed to operate at a moderate group index, i.e., optimizing equation~\eqref{eq:F1} and then equation~\eqref{eq:F2A}, are summarized in figure~\ref{fig: NDBP bands}. 
The band structure is shown in figure~\ref{fig: NDBP bands}(a). The NT mode has a SM slow light region with group index $n_g \approx 30$ deep inside the bandgap and away from the Brillouin zone edge.  At a frequency where $n_g=30$, indicated with a circular marker in figure~\ref{fig: NDBP bands}(a) and (b), the NT mode has $\max(\tilde{\PF}^f) = 4.5$, with $C=0.91$. In the region at least 35~nm from the nearest hole edge where $C \geq 0.95$, the maximum forwards Purcell factor is $\max(\PF^f) = 4.7$. This design satisfies our primary objective, the maximum Purcell factor at a region of high directionality far from a hole edge has been improved by a magnitude of order to values around our desired value of 5~\cite{PhysRevB.98.045309,Somaschi2016,PhysRevLett.116.020401}.

\begin{figure}[htb!]
    \centering
    \includegraphics[width=\linewidth]{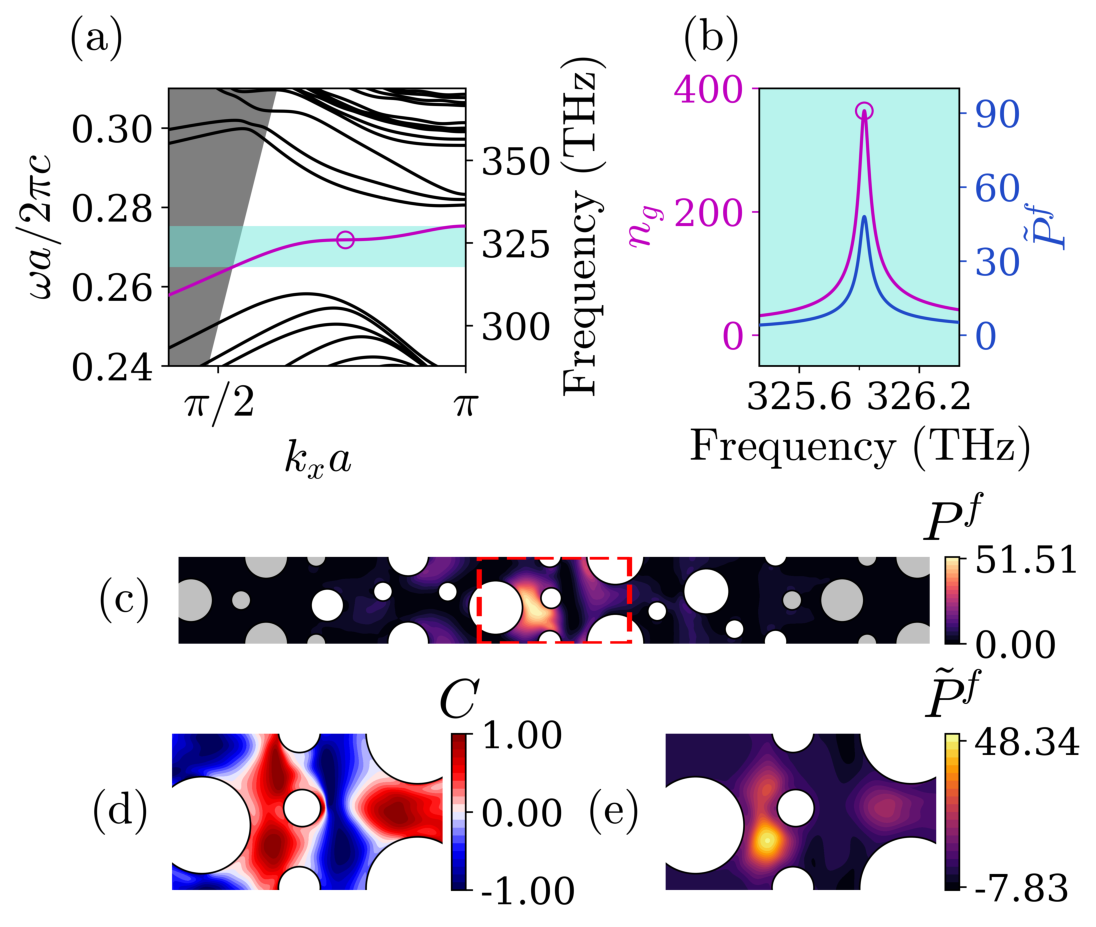}
    \caption{
    Summary of new design B.
    (a) Band structure. (b) Group index and the maximum chiral forwards Purcell factor for the NT mode. The following fields are calculated at the circular marker in (a) and (b) where $n_g = 360$. (c) Forwards Purcell factor $\PF^f$ for $\bm \mu = \bm \sigma_-$, (d) the associated chiral directionality $C$, and (e) the associated chiral forwards Purcell factor $\tilde{\PF^f}$. The holes which were not allowed to be modified during the optimization are drawn in grey while the holes which were are drawn in white.
    }
    \label{fig: allSMng bands}
\end{figure}

Next, we optimize a waveguide to operate in the slow-light regime, well away from the band edge, optimizing equation~\eqref{eq:F1} and then equation~\eqref{eq:F2B}. The results are summarized in figure~\ref{fig: allSMng bands}. For this design, we observe a slow light, SM bandwidth with a maximum $n_g > 350$ , deep inside the bandgap and away from the mode edge (Figs.~\ref{fig: allSMng bands}a and b). For the mode with $n_g = 360$, we find $\max(\tilde{\PF}^f) = 48$, with $C=0.98$. The position of $\max(\tilde{\PF}^f)$ is 40~nm away from the nearest hole edge. This is a two magnitude of order increase to the maximum chiral forwards Purcell factor from the initial design. 
In the region at least 35~nm from the nearest hole edge where $C \geq 0.95$, the maximum forwards Purcell factor is $\max(\PF^f) = 50$.

\added{In both of the improved designs, the glide plane symmetry of a reflection about the $x$ axis and translation in the $x$ direction of $a/2$ has been broken. This has served to increase the overlap of the region of circular polarization with areas of high optical energy in the unit cell therefore increasing the chiral forwards Purcell factor. In Ref.~\cite{hauffChiralQuantumOptics}, it was
also found that topological PCWs can outperform those which have been specifically designed for chiral coupling, namely the glide plane waveguide.}

\begin{comment}
\begin{table}
\centering
\caption{
Summary of designs: maximum chiral forwards Purcell factor $\tilde{\PF^f}$ at selected group index $n_g$ and the associated maximum forwards Purcell factor $\PF^f$ in a region with $C \geq 0.95$ at least 35~nm from the nearest hole edge.
}
\label{table: summary}
\begin{tabular}{c c c c} \toprule
     & $n_g$ & $\tilde{\PF^f}$ & $\PF^f$ \\ \midrule
	Initial design & 8.5 & 0.41 & 0.41 \\
	Design A & 30 & 4.5 & 4.7 \\
	Design B & 360 & 48 & 50 \\\bottomrule
\end{tabular}
\end{table}
\end{comment}

\section{Conclusions}
\label{sec:Conclusions} 

Using an efficient nanophotonic inverse design approach, we have optimized a state of the art topological PCWs for directional QD-photon interactions. We present PCWs designed to operate in the single mode regime, well away from the mode edge, and with moderate (30) and large (360) group indices. The designs have maximum forwards Purcell factors, at least 35~nm away from the nearest hole edge with $C >= 0.95$, of 4.7 and 50 at group indexes of 30 and 360 respectively.

Our optimized designs, while respecting all fabrication requirements, break the symmetry of the initial~\cite{mehrabadChiralTopologicalPhotonics2020} and other previously reported~\cite{hauffChiralQuantumOptics, heSilicononinsulatorSlabTopological2019, shalaevRobustTopologicallyProtected2019, yoshimiExperimentalDemonstrationTopological2021, barikTopologicalQuantumOptics2018} topological PCWs, as the first 3 original crystal unit cells on each side of interface differ (see Appendix B for details). As a result, the T mode disappears, ensuring the SM operation of the remaining, NT mode. Moreover, this broken symmetry leads to an asymmetric (with respect to the direction of propagation, $\hat{x}$) field (and hence, $\PF^f$) distributions. This is most noticeable for slow-light modes (cf.~Fig.~\ref{fig: allSMng bands}c) but can also be observed in the faster mode (cf.~Fig.~\ref{fig: NDBP bands}c). In essence, our optimization further increases the chiral-projected local density of  states  by {\it borrowing} from other regions, in a way that is not possible with standard, topological PCW designs. Our results demonstrate the power of efficient inverse optimization methods, providing a key to unlock the full potential of topological quantum photonics. 

\acknowledgements
We thank Nicholas Martin, Mahmoud Jalali Mehrabad, Andrew Foster, and Luke Wilson for useful discussions.
This work was supported by the Natural Sciences and Engineering Research Council of Canada, the Canadian Foundation for Innovation, and Queen's University, Canada.

%\pagebreak

%\clearpage
%\newpage
\vspace{1cm}

\appendix

\section{Additional edge states existing at the edges of the GME supercell}
\label{sec: AppA}

\begin{figure}[h]
    \centering
    \includegraphics[width=\linewidth]{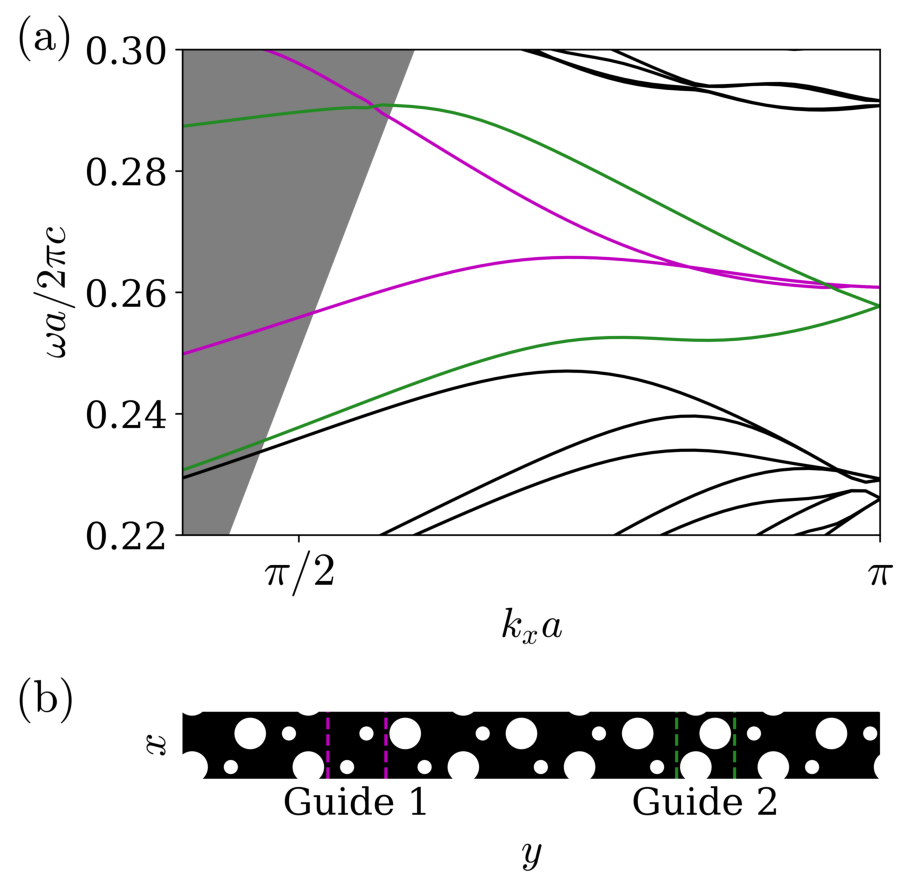}
    \caption{(a) Band structure for the unit cell in (b). The modes existing at guide 1 are drawn in purple, the modes existing at guide 2 are drawn in green, and the bulk modes are drawn in black. (b) GME unit cell containing the waveguide which the optimization was started from~\cite{mehrabadChiralTopologicalPhotonics2020} and another waveguide~\cite{heSilicononinsulatorSlabTopological2019}.}
    \label{fig: secondary bands}
\end{figure}

The GME method assumes 2D periodicity. As the crystals on each side of the waveguide are inversion symmetry partners, at the $y$-limits of the GME waveguide unit cell a second topological waveguide is modeled. This is a legitimate design, though not the design of interest. As an example the band structure for the unit cell shown in figure~\ref{fig: secondary bands}(b) is plotted in figure~\ref{fig: secondary bands}(a). The unit cell contains two interfaces between the topologically distinct crystals, the first being the waveguide~\cite{mehrabadChiralTopologicalPhotonics2020} which the optimization was started from and the second being another topological waveguide which has previously been described in Ref.~\cite{heSilicononinsulatorSlabTopological2019}.

\FloatBarrier
\section{Specifications of improved designs}

In this appendix, the position and radii of holes in the initial and improved designs are provided. Table~\ref{table: all holes inner} lists the position and radii of the 12 holes closest to the interface in the initial design, which are the holes whose position and radii were allowed to be modified during the optimizations. Table~\ref{table: all holes inner change} lists the change in position and radii of these holes for the new designs A and B.

% One big table vertical
\begin{table}
\centering
\caption{Location and radius of the 12 holes, indexed by $m$, closest to the interface in the initial design. The x axis is at the center of the waveguide initial design, see figure~\ref{fig: berry} for the coordinate system.
}
\label{table: all holes inner}
\begin{tabular}{c c c c c} \toprule
    Design & $m$ & $x [a]$ & $y [a]$ & $r [a]$\\ \midrule
    \multirow{12}{4em}{Initial Design} 	& 1 & 0.0000 & -2.4537 & 0.2350 \\
	 & 2 & 0.0000 & -1.8764 & 0.1053 \\
	 & 3 & 0.5000 & -1.5877 & 0.2350 \\
	 & 4 & 0.5000 & -1.0104 & 0.1053 \\
	 & 5 & 0.0000 & -0.7217 & 0.2350 \\
	 & 6 & 0.0000 & -0.1443 & 0.1053 \\
	 & 7 & 0.5000 & 0.1443 & 0.1053 \\
	 & 8 & 0.5000 & 0.7217 & 0.2350 \\
	 & 9 & 0.0000 & 1.0104 & 0.1053 \\
	 & 10 & 0.0000 & 1.5877 & 0.2350 \\
	 & 11 & 0.5000 & 1.8764 & 0.1053 \\
	 & 12 & 0.5000 & 2.4537 & 0.2350 \\\bottomrule
\end{tabular}
\end{table}

\begin{table}
\centering
\caption{Change in the location and radius of the 12 holes closest to the interface in design A and B. The hole indexes $m$ are the same as in table~\ref{table: all holes inner}.
}
\label{table: all holes inner change}
\begin{tabular}{c c c c c} \toprule
    Design & $m$ & $\Delta x [a]$ & $\Delta y [a]$ & $\Delta r [a]$\\ \midrule
    \multirow{12}{5em}{Design A} & 1 & 0.0316 & -0.1168 & -0.0349 \\
	 & 2 & -0.1319 & -0.2374 & -0.0020 \\
	 & 3 & -0.0121 & -0.1065 & 0.0461 \\
	 & 4 & 0.4557 & -0.2218 & -0.0020 \\
	 & 5 & 0.0089 & 0.0994 & 0.0922 \\
	 & 6 & -0.0059 & 0.2224 & -0.0020 \\
	 & 7 & -0.0078 & -0.2666 & 0.0699 \\
	 & 8 & -0.0134 & -0.0060 & 0.0760 \\
	 & 9 & 0.1667 & 0.2083 & -0.0020 \\
	 & 10 & -0.1048 & 0.1439 & 0.0598 \\
	 & 11 & -0.1484 & 0.2485 & -0.0020 \\
	 & 12 & 0.0047 & 0.1063 & -0.0591 \\\midrule
    \multirow{12}{4em}{Design B} & 1 & 0.0579 & -0.1602 & -0.0616 \\
	 & 2 & -0.1001 & -0.0970 & -0.0020 \\
	 & 3 & -0.0150 & -0.0948 & -0.0133 \\
	 & 4 & 0.4028 & -0.2156 & -0.0020 \\
	 & 5 & 0.0849 & 0.0470 & 0.0606 \\
	 & 6 & -0.0244 & 0.1112 & 0.0068 \\
	 & 7 & -0.0146 & -0.1952 & 0.0227 \\
	 & 8 & -0.0101 & -0.0157 & 0.0801 \\
	 & 9 & 0.1269 & 0.1780 & -0.0020 \\
	 & 10 & -0.1016 & 0.1668 & 0.0126 \\
	 & 11 & -0.1662 & 0.2031 & -0.0020 \\
	 & 12 & -0.0236 & 0.0983 & -0.1173 \\\bottomrule
\end{tabular}
\end{table}

\clearpage
\newpage
\FloatBarrier
\bibliography{refs}

\begin{thebibliography}{64}%
\makeatletter
\providecommand \@ifxundefined [1]{%
 \@ifx{#1\undefined}
}%
\providecommand \@ifnum [1]{%
 \ifnum #1\expandafter \@firstoftwo
 \else \expandafter \@secondoftwo
 \fi
}%
\providecommand \@ifx [1]{%
 \ifx #1\expandafter \@firstoftwo
 \else \expandafter \@secondoftwo
 \fi
}%
\providecommand \natexlab [1]{#1}%
\providecommand \enquote  [1]{``#1''}%
\providecommand \bibnamefont  [1]{#1}%
\providecommand \bibfnamefont [1]{#1}%
\providecommand \citenamefont [1]{#1}%
\providecommand \href@noop [0]{\@secondoftwo}%
\providecommand \href [0]{\begingroup \@sanitize@url \@href}%
\providecommand \@href[1]{\@@startlink{#1}\@@href}%
\providecommand \@@href[1]{\endgroup#1\@@endlink}%
\providecommand \@sanitize@url [0]{\catcode `\\12\catcode `\$12\catcode
  `\&12\catcode `\#12\catcode `\^12\catcode `\_12\catcode `\%12\relax}%
\providecommand \@@startlink[1]{}%
\providecommand \@@endlink[0]{}%
\providecommand \url  [0]{\begingroup\@sanitize@url \@url }%
\providecommand \@url [1]{\endgroup\@href {#1}{\urlprefix }}%
\providecommand \urlprefix  [0]{URL }%
\providecommand \Eprint [0]{\href }%
\providecommand \doibase [0]{https://doi.org/}%
\providecommand \selectlanguage [0]{\@gobble}%
\providecommand \bibinfo  [0]{\@secondoftwo}%
\providecommand \bibfield  [0]{\@secondoftwo}%
\providecommand \translation [1]{[#1]}%
\providecommand \BibitemOpen [0]{}%
\providecommand \bibitemStop [0]{}%
\providecommand \bibitemNoStop [0]{.\EOS\space}%
\providecommand \EOS [0]{\spacefactor3000\relax}%
\providecommand \BibitemShut  [1]{\csname bibitem#1\endcsname}%
\let\auto@bib@innerbib\@empty
%</preamble>
\bibitem [{\citenamefont {Lodahl}\ \emph {et~al.}(2017)\citenamefont {Lodahl},
  \citenamefont {Mahmoodian}, \citenamefont {Stobbe}, \citenamefont
  {Rauschenbeutel}, \citenamefont {Schneeweiss}, \citenamefont {Volz},
  \citenamefont {Pichler},\ and\ \citenamefont {Zoller}}]{Lodahl2017}%
  \BibitemOpen
  \bibfield  {author} {\bibinfo {author} {\bibfnamefont {P.}~\bibnamefont
  {Lodahl}}, \bibinfo {author} {\bibfnamefont {S.}~\bibnamefont {Mahmoodian}},
  \bibinfo {author} {\bibfnamefont {S.}~\bibnamefont {Stobbe}}, \bibinfo
  {author} {\bibfnamefont {A.}~\bibnamefont {Rauschenbeutel}}, \bibinfo
  {author} {\bibfnamefont {P.}~\bibnamefont {Schneeweiss}}, \bibinfo {author}
  {\bibfnamefont {J.}~\bibnamefont {Volz}}, \bibinfo {author} {\bibfnamefont
  {H.}~\bibnamefont {Pichler}},\ and\ \bibinfo {author} {\bibfnamefont
  {P.}~\bibnamefont {Zoller}},\ }\bibfield  {title} {\bibinfo {title} {Chiral
  quantum optics},\ }\href {https://doi.org/10.1038/nature21037} {\bibfield
  {journal} {\bibinfo  {journal} {Nature}\ }\textbf {\bibinfo {volume} {541}},\
  \bibinfo {pages} {473} (\bibinfo {year} {2017})}\BibitemShut {NoStop}%
\bibitem [{\citenamefont {S{\"o}llner}\ \emph {et~al.}(2015)\citenamefont
  {S{\"o}llner}, \citenamefont {Mahmoodian}, \citenamefont {Hansen},
  \citenamefont {Midolo}, \citenamefont {Javadi}, \citenamefont {Kir{\v
  s}ansk{\.e}}, \citenamefont {Pregnolato}, \citenamefont {{El-Ella}},
  \citenamefont {Lee}, \citenamefont {Song}, \citenamefont {Stobbe},\ and\
  \citenamefont {Lodahl}}]{sollnerDeterministicPhotonEmitter2015}%
  \BibitemOpen
  \bibfield  {author} {\bibinfo {author} {\bibfnamefont {I.}~\bibnamefont
  {S{\"o}llner}}, \bibinfo {author} {\bibfnamefont {S.}~\bibnamefont
  {Mahmoodian}}, \bibinfo {author} {\bibfnamefont {S.~L.}\ \bibnamefont
  {Hansen}}, \bibinfo {author} {\bibfnamefont {L.}~\bibnamefont {Midolo}},
  \bibinfo {author} {\bibfnamefont {A.}~\bibnamefont {Javadi}}, \bibinfo
  {author} {\bibfnamefont {G.}~\bibnamefont {Kir{\v s}ansk{\.e}}}, \bibinfo
  {author} {\bibfnamefont {T.}~\bibnamefont {Pregnolato}}, \bibinfo {author}
  {\bibfnamefont {H.}~\bibnamefont {{El-Ella}}}, \bibinfo {author}
  {\bibfnamefont {E.~H.}\ \bibnamefont {Lee}}, \bibinfo {author} {\bibfnamefont
  {J.~D.}\ \bibnamefont {Song}}, \bibinfo {author} {\bibfnamefont
  {S.}~\bibnamefont {Stobbe}},\ and\ \bibinfo {author} {\bibfnamefont
  {P.}~\bibnamefont {Lodahl}},\ }\bibfield  {title} {\bibinfo {title}
  {Deterministic photon\textendash emitter coupling in chiral photonic
  circuits},\ }\href {https://doi.org/10.1038/nnano.2015.159} {\bibfield
  {journal} {\bibinfo  {journal} {Nature Nanotechnology}\ }\textbf {\bibinfo
  {volume} {10}},\ \bibinfo {pages} {775} (\bibinfo {year} {2015})}\BibitemShut
  {NoStop}%
\bibitem [{\citenamefont {Xia}\ \emph {et~al.}(2014)\citenamefont {Xia},
  \citenamefont {Lu}, \citenamefont {Lin}, \citenamefont {Cheng}, \citenamefont
  {Niu}, \citenamefont {Gong},\ and\ \citenamefont
  {Twamley}}]{PhysRevA.90.043802}%
  \BibitemOpen
  \bibfield  {author} {\bibinfo {author} {\bibfnamefont {K.}~\bibnamefont
  {Xia}}, \bibinfo {author} {\bibfnamefont {G.}~\bibnamefont {Lu}}, \bibinfo
  {author} {\bibfnamefont {G.}~\bibnamefont {Lin}}, \bibinfo {author}
  {\bibfnamefont {Y.}~\bibnamefont {Cheng}}, \bibinfo {author} {\bibfnamefont
  {Y.}~\bibnamefont {Niu}}, \bibinfo {author} {\bibfnamefont {S.}~\bibnamefont
  {Gong}},\ and\ \bibinfo {author} {\bibfnamefont {J.}~\bibnamefont
  {Twamley}},\ }\bibfield  {title} {\bibinfo {title} {Reversible nonmagnetic
  single-photon isolation using unbalanced quantum coupling},\ }\href
  {https://doi.org/10.1103/PhysRevA.90.043802} {\bibfield  {journal} {\bibinfo
  {journal} {Phys. Rev. A}\ }\textbf {\bibinfo {volume} {90}},\ \bibinfo
  {pages} {043802} (\bibinfo {year} {2014})}\BibitemShut {NoStop}%
\bibitem [{\citenamefont {Sayrin}\ \emph {et~al.}(2015)\citenamefont {Sayrin},
  \citenamefont {Junge}, \citenamefont {Mitsch}, \citenamefont {Albrecht},
  \citenamefont {O'Shea}, \citenamefont {Schneeweiss}, \citenamefont {Volz},\
  and\ \citenamefont {Rauschenbeutel}}]{PhysRevX.5.041036}%
  \BibitemOpen
  \bibfield  {author} {\bibinfo {author} {\bibfnamefont {C.}~\bibnamefont
  {Sayrin}}, \bibinfo {author} {\bibfnamefont {C.}~\bibnamefont {Junge}},
  \bibinfo {author} {\bibfnamefont {R.}~\bibnamefont {Mitsch}}, \bibinfo
  {author} {\bibfnamefont {B.}~\bibnamefont {Albrecht}}, \bibinfo {author}
  {\bibfnamefont {D.}~\bibnamefont {O'Shea}}, \bibinfo {author} {\bibfnamefont
  {P.}~\bibnamefont {Schneeweiss}}, \bibinfo {author} {\bibfnamefont
  {J.}~\bibnamefont {Volz}},\ and\ \bibinfo {author} {\bibfnamefont
  {A.}~\bibnamefont {Rauschenbeutel}},\ }\bibfield  {title} {\bibinfo {title}
  {Nanophotonic optical isolator controlled by the internal state of cold
  atoms},\ }\href {https://doi.org/10.1103/PhysRevX.5.041036} {\bibfield
  {journal} {\bibinfo  {journal} {Phys. Rev. X}\ }\textbf {\bibinfo {volume}
  {5}},\ \bibinfo {pages} {041036} (\bibinfo {year} {2015})}\BibitemShut
  {NoStop}%
\bibitem [{\citenamefont {Scheucher}\ \emph {et~al.}(2016)\citenamefont
  {Scheucher}, \citenamefont {Hilico}, \citenamefont {Will}, \citenamefont
  {Volz},\ and\ \citenamefont {Rauschenbeutel}}]{Scheucher2016}%
  \BibitemOpen
  \bibfield  {author} {\bibinfo {author} {\bibfnamefont {M.}~\bibnamefont
  {Scheucher}}, \bibinfo {author} {\bibfnamefont {A.}~\bibnamefont {Hilico}},
  \bibinfo {author} {\bibfnamefont {E.}~\bibnamefont {Will}}, \bibinfo {author}
  {\bibfnamefont {J.}~\bibnamefont {Volz}},\ and\ \bibinfo {author}
  {\bibfnamefont {A.}~\bibnamefont {Rauschenbeutel}},\ }\bibfield  {title}
  {\bibinfo {title} {Quantum optical circulator controlled by a single chirally
  coupled atom},\ }\href {https://doi.org/10.1126/science.aaj2118} {\bibfield
  {journal} {\bibinfo  {journal} {Science}\ }\textbf {\bibinfo {volume}
  {354}},\ \bibinfo {pages} {1577} (\bibinfo {year} {2016})}\BibitemShut
  {NoStop}%
\bibitem [{\citenamefont {Pucher}\ \emph {et~al.}(2022)\citenamefont {Pucher},
  \citenamefont {Liedl}, \citenamefont {Jin}, \citenamefont {Rauschenbeutel},\
  and\ \citenamefont
  {Schneeweiss}}]{pucherAtomicSpincontrolledNonreciprocal2022}%
  \BibitemOpen
  \bibfield  {author} {\bibinfo {author} {\bibfnamefont {S.}~\bibnamefont
  {Pucher}}, \bibinfo {author} {\bibfnamefont {C.}~\bibnamefont {Liedl}},
  \bibinfo {author} {\bibfnamefont {S.}~\bibnamefont {Jin}}, \bibinfo {author}
  {\bibfnamefont {A.}~\bibnamefont {Rauschenbeutel}},\ and\ \bibinfo {author}
  {\bibfnamefont {P.}~\bibnamefont {Schneeweiss}},\ }\bibfield  {title}
  {\bibinfo {title} {Atomic spin-controlled non-reciprocal {{Raman}}
  amplification of fibre-guided light},\ }\href
  {https://doi.org/10.1038/s41566-022-00987-z} {\bibfield  {journal} {\bibinfo
  {journal} {Nature Photonics}\ }\textbf {\bibinfo {volume} {16}},\ \bibinfo
  {pages} {380} (\bibinfo {year} {2022})}\BibitemShut {NoStop}%
\bibitem [{\citenamefont {Shomroni}\ \emph {et~al.}(2014)\citenamefont
  {Shomroni}, \citenamefont {Rosenblum}, \citenamefont {Lovsky}, \citenamefont
  {Bechler}, \citenamefont {Guendelman},\ and\ \citenamefont
  {Dayan}}]{Shomroni2014}%
  \BibitemOpen
  \bibfield  {author} {\bibinfo {author} {\bibfnamefont {I.}~\bibnamefont
  {Shomroni}}, \bibinfo {author} {\bibfnamefont {S.}~\bibnamefont {Rosenblum}},
  \bibinfo {author} {\bibfnamefont {Y.}~\bibnamefont {Lovsky}}, \bibinfo
  {author} {\bibfnamefont {O.}~\bibnamefont {Bechler}}, \bibinfo {author}
  {\bibfnamefont {G.}~\bibnamefont {Guendelman}},\ and\ \bibinfo {author}
  {\bibfnamefont {B.}~\bibnamefont {Dayan}},\ }\bibfield  {title} {\bibinfo
  {title} {All-optical routing of single photons by a one-atom switch
  controlled by a single photon},\ }\href
  {https://doi.org/10.1126/science.1254699} {\bibfield  {journal} {\bibinfo
  {journal} {Science}\ }\textbf {\bibinfo {volume} {345}},\ \bibinfo {pages}
  {903} (\bibinfo {year} {2014})}\BibitemShut {NoStop}%
\bibitem [{\citenamefont {Østfeldt}\ \emph {et~al.}(2021)\citenamefont
  {Østfeldt}, \citenamefont {González-Ruiz}, \citenamefont {Hauff},
  \citenamefont {Wang}, \citenamefont {Wieck}, \citenamefont {Ludwig},
  \citenamefont {Schott}, \citenamefont {Midolo}, \citenamefont {Sørensen},
  \citenamefont {Uppu},\ and\ \citenamefont {Lodahl}}]{2109.03519}%
  \BibitemOpen
  \bibfield  {author} {\bibinfo {author} {\bibfnamefont {F.~T.}\ \bibnamefont
  {Østfeldt}}, \bibinfo {author} {\bibfnamefont {E.~M.}\ \bibnamefont
  {González-Ruiz}}, \bibinfo {author} {\bibfnamefont {N.}~\bibnamefont
  {Hauff}}, \bibinfo {author} {\bibfnamefont {Y.}~\bibnamefont {Wang}},
  \bibinfo {author} {\bibfnamefont {A.~D.}\ \bibnamefont {Wieck}}, \bibinfo
  {author} {\bibfnamefont {A.}~\bibnamefont {Ludwig}}, \bibinfo {author}
  {\bibfnamefont {R.}~\bibnamefont {Schott}}, \bibinfo {author} {\bibfnamefont
  {L.}~\bibnamefont {Midolo}}, \bibinfo {author} {\bibfnamefont {A.~S.}\
  \bibnamefont {Sørensen}}, \bibinfo {author} {\bibfnamefont {R.}~\bibnamefont
  {Uppu}},\ and\ \bibinfo {author} {\bibfnamefont {P.}~\bibnamefont {Lodahl}},\
  }\href@noop {} {\bibinfo {title} {On-demand source of dual-rail photon pairs
  based on chiral interaction in a nanophotonic waveguide}} (\bibinfo {year}
  {2021}),\ \Eprint {https://arxiv.org/abs/arXiv:2109.03519} {arXiv:2109.03519}
  \BibitemShut {NoStop}%
\bibitem [{\citenamefont {Mahmoodian}\ \emph {et~al.}(2016)\citenamefont
  {Mahmoodian}, \citenamefont {Lodahl},\ and\ \citenamefont
  {S\o{}rensen}}]{PhysRevLett.117.240501}%
  \BibitemOpen
  \bibfield  {author} {\bibinfo {author} {\bibfnamefont {S.}~\bibnamefont
  {Mahmoodian}}, \bibinfo {author} {\bibfnamefont {P.}~\bibnamefont {Lodahl}},\
  and\ \bibinfo {author} {\bibfnamefont {A.~S.}\ \bibnamefont {S\o{}rensen}},\
  }\bibfield  {title} {\bibinfo {title} {Quantum networks with
  chiral-light--matter interaction in waveguides},\ }\href
  {https://doi.org/10.1103/PhysRevLett.117.240501} {\bibfield  {journal}
  {\bibinfo  {journal} {Phys. Rev. Lett.}\ }\textbf {\bibinfo {volume} {117}},\
  \bibinfo {pages} {240501} (\bibinfo {year} {2016})}\BibitemShut {NoStop}%
\bibitem [{\citenamefont {Manga~Rao}\ and\ \citenamefont
  {Hughes}(2007)}]{mangaraoSingleQuantumdotPurcell2007}%
  \BibitemOpen
  \bibfield  {author} {\bibinfo {author} {\bibfnamefont {V.~S.~C.}\
  \bibnamefont {Manga~Rao}}\ and\ \bibinfo {author} {\bibfnamefont
  {S.}~\bibnamefont {Hughes}},\ }\bibfield  {title} {\bibinfo {title} {Single
  quantum-dot {{Purcell}} factor and {$\beta$} factor in a photonic crystal
  waveguide},\ }\href {https://doi.org/10.1103/PhysRevB.75.205437} {\bibfield
  {journal} {\bibinfo  {journal} {Physical Review B}\ }\textbf {\bibinfo
  {volume} {75}},\ \bibinfo {pages} {205437} (\bibinfo {year}
  {2007})}\BibitemShut {NoStop}%
\bibitem [{\citenamefont {Laucht}\ \emph {et~al.}(2012)\citenamefont {Laucht},
  \citenamefont {P{\"u}tz}, \citenamefont {G{\"u}nthner}, \citenamefont
  {Hauke}, \citenamefont {Saive}, \citenamefont {Fr{\'e}d{\'e}rick},
  \citenamefont {Bichler}, \citenamefont {Amann}, \citenamefont {Holleitner},
  \citenamefont {Kaniber},\ and\ \citenamefont
  {Finley}}]{lauchtWaveguideCoupledOnChipSinglePhoton2012}%
  \BibitemOpen
  \bibfield  {author} {\bibinfo {author} {\bibfnamefont {A.}~\bibnamefont
  {Laucht}}, \bibinfo {author} {\bibfnamefont {S.}~\bibnamefont {P{\"u}tz}},
  \bibinfo {author} {\bibfnamefont {T.}~\bibnamefont {G{\"u}nthner}}, \bibinfo
  {author} {\bibfnamefont {N.}~\bibnamefont {Hauke}}, \bibinfo {author}
  {\bibfnamefont {R.}~\bibnamefont {Saive}}, \bibinfo {author} {\bibfnamefont
  {S.}~\bibnamefont {Fr{\'e}d{\'e}rick}}, \bibinfo {author} {\bibfnamefont
  {M.}~\bibnamefont {Bichler}}, \bibinfo {author} {\bibfnamefont {M.-C.}\
  \bibnamefont {Amann}}, \bibinfo {author} {\bibfnamefont {A.~W.}\ \bibnamefont
  {Holleitner}}, \bibinfo {author} {\bibfnamefont {M.}~\bibnamefont
  {Kaniber}},\ and\ \bibinfo {author} {\bibfnamefont {J.~J.}\ \bibnamefont
  {Finley}},\ }\bibfield  {title} {\bibinfo {title} {A {{Waveguide}}-{{Coupled
  On}}-{{Chip Single}}-{{Photon Source}}},\ }\href
  {https://doi.org/10.1103/PhysRevX.2.011014} {\bibfield  {journal} {\bibinfo
  {journal} {Physical Review X}\ }\textbf {\bibinfo {volume} {2}},\ \bibinfo
  {pages} {011014} (\bibinfo {year} {2012})}\BibitemShut {NoStop}%
\bibitem [{\citenamefont {Lodahl}\ \emph
  {et~al.}(2015{\natexlab{a}})\citenamefont {Lodahl}, \citenamefont
  {Mahmoodian},\ and\ \citenamefont {Stobbe}}]{RevModPhys.87.347}%
  \BibitemOpen
  \bibfield  {author} {\bibinfo {author} {\bibfnamefont {P.}~\bibnamefont
  {Lodahl}}, \bibinfo {author} {\bibfnamefont {S.}~\bibnamefont {Mahmoodian}},\
  and\ \bibinfo {author} {\bibfnamefont {S.}~\bibnamefont {Stobbe}},\
  }\bibfield  {title} {\bibinfo {title} {Interfacing single photons and single
  quantum dots with photonic nanostructures},\ }\href
  {https://doi.org/10.1103/RevModPhys.87.347} {\bibfield  {journal} {\bibinfo
  {journal} {Rev. Mod. Phys.}\ }\textbf {\bibinfo {volume} {87}},\ \bibinfo
  {pages} {347} (\bibinfo {year} {2015}{\natexlab{a}})}\BibitemShut {NoStop}%
\bibitem [{\citenamefont {Kuruma}\ \emph {et~al.}()\citenamefont {Kuruma},
  \citenamefont {Yoshimi}, \citenamefont {Ota}, \citenamefont {Katsumi},
  \citenamefont {Kakuda}, \citenamefont {Arakawa},\ and\ \citenamefont
  {Iwamoto}}]{kurumaTopologicallyProtectedSinglePhotonSources}%
  \BibitemOpen
  \bibfield  {author} {\bibinfo {author} {\bibfnamefont {K.}~\bibnamefont
  {Kuruma}}, \bibinfo {author} {\bibfnamefont {H.}~\bibnamefont {Yoshimi}},
  \bibinfo {author} {\bibfnamefont {Y.}~\bibnamefont {Ota}}, \bibinfo {author}
  {\bibfnamefont {R.}~\bibnamefont {Katsumi}}, \bibinfo {author} {\bibfnamefont
  {M.}~\bibnamefont {Kakuda}}, \bibinfo {author} {\bibfnamefont
  {Y.}~\bibnamefont {Arakawa}},\ and\ \bibinfo {author} {\bibfnamefont
  {S.}~\bibnamefont {Iwamoto}},\ }\bibfield  {title} {\bibinfo {title}
  {Topologically-{{Protected Single-Photon Sources}} with {{Topological Slow
  Light Photonic Crystal Waveguides}}},\ }\href
  {https://doi.org/10.1002/lpor.202200077} {\bibfield  {journal} {\bibinfo
  {journal} {Laser \& Photonics Reviews}\ }\textbf {\bibinfo {volume} {n/a}},\
  \bibinfo {pages} {2200077}}\BibitemShut {NoStop}%
\bibitem [{\citenamefont {Arcari}\ \emph {et~al.}(2014)\citenamefont {Arcari},
  \citenamefont {S{\"o}llner}, \citenamefont {Javadi}, \citenamefont
  {Lindskov~Hansen}, \citenamefont {Mahmoodian}, \citenamefont {Liu},
  \citenamefont {Thyrrestrup}, \citenamefont {Lee}, \citenamefont {Song},
  \citenamefont {Stobbe},\ and\ \citenamefont
  {Lodahl}}]{arcariNearUnityCouplingEfficiency2014}%
  \BibitemOpen
  \bibfield  {author} {\bibinfo {author} {\bibfnamefont {M.}~\bibnamefont
  {Arcari}}, \bibinfo {author} {\bibfnamefont {I.}~\bibnamefont {S{\"o}llner}},
  \bibinfo {author} {\bibfnamefont {A.}~\bibnamefont {Javadi}}, \bibinfo
  {author} {\bibfnamefont {S.}~\bibnamefont {Lindskov~Hansen}}, \bibinfo
  {author} {\bibfnamefont {S.}~\bibnamefont {Mahmoodian}}, \bibinfo {author}
  {\bibfnamefont {J.}~\bibnamefont {Liu}}, \bibinfo {author} {\bibfnamefont
  {H.}~\bibnamefont {Thyrrestrup}}, \bibinfo {author} {\bibfnamefont {E.~H.}\
  \bibnamefont {Lee}}, \bibinfo {author} {\bibfnamefont {J.~D.}\ \bibnamefont
  {Song}}, \bibinfo {author} {\bibfnamefont {S.}~\bibnamefont {Stobbe}},\ and\
  \bibinfo {author} {\bibfnamefont {P.}~\bibnamefont {Lodahl}},\ }\bibfield
  {title} {\bibinfo {title} {Near-{{Unity Coupling Efficiency}} of a {{Quantum
  Emitter}} to a {{Photonic Crystal Waveguide}}},\ }\href
  {https://doi.org/10.1103/PhysRevLett.113.093603} {\bibfield  {journal}
  {\bibinfo  {journal} {Physical Review Letters}\ }\textbf {\bibinfo {volume}
  {113}},\ \bibinfo {pages} {093603} (\bibinfo {year} {2014})}\BibitemShut
  {NoStop}%
\bibitem [{\citenamefont {Holz}\ \emph {et~al.}(2020)\citenamefont {Holz},
  \citenamefont {Auchter}, \citenamefont {Stocker}, \citenamefont {Valentini},
  \citenamefont {Lakhmanskiy}, \citenamefont {R\"{o}ssler}, \citenamefont
  {Stampfer}, \citenamefont {Sgouridis}, \citenamefont {Aschauer},
  \citenamefont {Colombe},\ and\ \citenamefont {Blatt}}]{Holz2020}%
  \BibitemOpen
  \bibfield  {author} {\bibinfo {author} {\bibfnamefont {P.~C.}\ \bibnamefont
  {Holz}}, \bibinfo {author} {\bibfnamefont {S.}~\bibnamefont {Auchter}},
  \bibinfo {author} {\bibfnamefont {G.}~\bibnamefont {Stocker}}, \bibinfo
  {author} {\bibfnamefont {M.}~\bibnamefont {Valentini}}, \bibinfo {author}
  {\bibfnamefont {K.}~\bibnamefont {Lakhmanskiy}}, \bibinfo {author}
  {\bibfnamefont {C.}~\bibnamefont {R\"{o}ssler}}, \bibinfo {author}
  {\bibfnamefont {P.}~\bibnamefont {Stampfer}}, \bibinfo {author}
  {\bibfnamefont {S.}~\bibnamefont {Sgouridis}}, \bibinfo {author}
  {\bibfnamefont {E.}~\bibnamefont {Aschauer}}, \bibinfo {author}
  {\bibfnamefont {Y.}~\bibnamefont {Colombe}},\ and\ \bibinfo {author}
  {\bibfnamefont {R.}~\bibnamefont {Blatt}},\ }\bibfield  {title} {\bibinfo
  {title} {2d linear trap array for quantum information processing},\ }\href
  {https://doi.org/10.1002/qute.202000031} {\bibfield  {journal} {\bibinfo
  {journal} {Advanced Quantum Technologies}\ }\textbf {\bibinfo {volume} {3}},\
  \bibinfo {pages} {2000031} (\bibinfo {year} {2020})}\BibitemShut {NoStop}%
\bibitem [{\citenamefont {Uppu}\ \emph {et~al.}(2020)\citenamefont {Uppu},
  \citenamefont {Pedersen}, \citenamefont {Wang}, \citenamefont {Olesen},
  \citenamefont {Papon}, \citenamefont {Zhou}, \citenamefont {Midolo},
  \citenamefont {Scholz}, \citenamefont {Wieck}, \citenamefont {Ludwig},\ and\
  \citenamefont {Lodahl}}]{Uppu2020}%
  \BibitemOpen
  \bibfield  {author} {\bibinfo {author} {\bibfnamefont {R.}~\bibnamefont
  {Uppu}}, \bibinfo {author} {\bibfnamefont {F.~T.}\ \bibnamefont {Pedersen}},
  \bibinfo {author} {\bibfnamefont {Y.}~\bibnamefont {Wang}}, \bibinfo {author}
  {\bibfnamefont {C.~T.}\ \bibnamefont {Olesen}}, \bibinfo {author}
  {\bibfnamefont {C.}~\bibnamefont {Papon}}, \bibinfo {author} {\bibfnamefont
  {X.}~\bibnamefont {Zhou}}, \bibinfo {author} {\bibfnamefont {L.}~\bibnamefont
  {Midolo}}, \bibinfo {author} {\bibfnamefont {S.}~\bibnamefont {Scholz}},
  \bibinfo {author} {\bibfnamefont {A.~D.}\ \bibnamefont {Wieck}}, \bibinfo
  {author} {\bibfnamefont {A.}~\bibnamefont {Ludwig}},\ and\ \bibinfo {author}
  {\bibfnamefont {P.}~\bibnamefont {Lodahl}},\ }\bibfield  {title} {\bibinfo
  {title} {Scalable integrated single-photon source},\ }\bibfield  {journal}
  {\bibinfo  {journal} {Science Advances}\ }\textbf {\bibinfo {volume} {6}},\
  \href {https://doi.org/10.1126/sciadv.abc8268} {10.1126/sciadv.abc8268}
  (\bibinfo {year} {2020})\BibitemShut {NoStop}%
\bibitem [{\citenamefont {Le~Jeannic}\ \emph {et~al.}(2021)\citenamefont
  {Le~Jeannic}, \citenamefont {Ramos}, \citenamefont {Simonsen}, \citenamefont
  {Pregnolato}, \citenamefont {Liu}, \citenamefont {Schott}, \citenamefont
  {Wieck}, \citenamefont {Ludwig}, \citenamefont {Rotenberg}, \citenamefont
  {Garc\'{\i}a-Ripoll},\ and\ \citenamefont {Lodahl}}]{PhysRevLett.126.023603}%
  \BibitemOpen
  \bibfield  {author} {\bibinfo {author} {\bibfnamefont {H.}~\bibnamefont
  {Le~Jeannic}}, \bibinfo {author} {\bibfnamefont {T.}~\bibnamefont {Ramos}},
  \bibinfo {author} {\bibfnamefont {S.~F.}\ \bibnamefont {Simonsen}}, \bibinfo
  {author} {\bibfnamefont {T.}~\bibnamefont {Pregnolato}}, \bibinfo {author}
  {\bibfnamefont {Z.}~\bibnamefont {Liu}}, \bibinfo {author} {\bibfnamefont
  {R.}~\bibnamefont {Schott}}, \bibinfo {author} {\bibfnamefont {A.~D.}\
  \bibnamefont {Wieck}}, \bibinfo {author} {\bibfnamefont {A.}~\bibnamefont
  {Ludwig}}, \bibinfo {author} {\bibfnamefont {N.}~\bibnamefont {Rotenberg}},
  \bibinfo {author} {\bibfnamefont {J.~J.}\ \bibnamefont
  {Garc\'{\i}a-Ripoll}},\ and\ \bibinfo {author} {\bibfnamefont
  {P.}~\bibnamefont {Lodahl}},\ }\bibfield  {title} {\bibinfo {title}
  {Experimental reconstruction of the few-photon nonlinear scattering matrix
  from a single quantum dot in a nanophotonic waveguide},\ }\href
  {https://doi.org/10.1103/PhysRevLett.126.023603} {\bibfield  {journal}
  {\bibinfo  {journal} {Phys. Rev. Lett.}\ }\textbf {\bibinfo {volume} {126}},\
  \bibinfo {pages} {023603} (\bibinfo {year} {2021})}\BibitemShut {NoStop}%
\bibitem [{\citenamefont {Lang}\ \emph {et~al.}(2015)\citenamefont {Lang},
  \citenamefont {Beggs}, \citenamefont {Young}, \citenamefont {Rarity},\ and\
  \citenamefont {Oulton}}]{PhysRevA.92.063819}%
  \BibitemOpen
  \bibfield  {author} {\bibinfo {author} {\bibfnamefont {B.}~\bibnamefont
  {Lang}}, \bibinfo {author} {\bibfnamefont {D.~M.}\ \bibnamefont {Beggs}},
  \bibinfo {author} {\bibfnamefont {A.~B.}\ \bibnamefont {Young}}, \bibinfo
  {author} {\bibfnamefont {J.~G.}\ \bibnamefont {Rarity}},\ and\ \bibinfo
  {author} {\bibfnamefont {R.}~\bibnamefont {Oulton}},\ }\bibfield  {title}
  {\bibinfo {title} {Stability of polarization singularities in disordered
  photonic crystal waveguides},\ }\href
  {https://doi.org/10.1103/PhysRevA.92.063819} {\bibfield  {journal} {\bibinfo
  {journal} {Phys. Rev. A}\ }\textbf {\bibinfo {volume} {92}},\ \bibinfo
  {pages} {063819} (\bibinfo {year} {2015})}\BibitemShut {NoStop}%
\bibitem [{\citenamefont {Lang}\ \emph {et~al.}(2022)\citenamefont {Lang},
  \citenamefont {McCutcheon}, \citenamefont {Harbord}, \citenamefont {Young},\
  and\ \citenamefont {Oulton}}]{PhysRevLett.128.073602}%
  \BibitemOpen
  \bibfield  {author} {\bibinfo {author} {\bibfnamefont {B.}~\bibnamefont
  {Lang}}, \bibinfo {author} {\bibfnamefont {D.~P.~S.}\ \bibnamefont
  {McCutcheon}}, \bibinfo {author} {\bibfnamefont {E.}~\bibnamefont {Harbord}},
  \bibinfo {author} {\bibfnamefont {A.~B.}\ \bibnamefont {Young}},\ and\
  \bibinfo {author} {\bibfnamefont {R.}~\bibnamefont {Oulton}},\ }\bibfield
  {title} {\bibinfo {title} {Perfect chirality with imperfect polarization},\
  }\href {https://doi.org/10.1103/PhysRevLett.128.073602} {\bibfield  {journal}
  {\bibinfo  {journal} {Phys. Rev. Lett.}\ }\textbf {\bibinfo {volume} {128}},\
  \bibinfo {pages} {073602} (\bibinfo {year} {2022})}\BibitemShut {NoStop}%
\bibitem [{\citenamefont {Young}\ \emph {et~al.}(2015)\citenamefont {Young},
  \citenamefont {Thijssen}, \citenamefont {Beggs}, \citenamefont
  {Androvitsaneas}, \citenamefont {Kuipers}, \citenamefont {Rarity},
  \citenamefont {Hughes},\ and\ \citenamefont
  {Oulton}}]{youngPolarizationEngineeringPhotonic2015}%
  \BibitemOpen
  \bibfield  {author} {\bibinfo {author} {\bibfnamefont {A.~B.}\ \bibnamefont
  {Young}}, \bibinfo {author} {\bibfnamefont {A.~C.~T.}\ \bibnamefont
  {Thijssen}}, \bibinfo {author} {\bibfnamefont {D.~M.}\ \bibnamefont {Beggs}},
  \bibinfo {author} {\bibfnamefont {P.}~\bibnamefont {Androvitsaneas}},
  \bibinfo {author} {\bibfnamefont {L.}~\bibnamefont {Kuipers}}, \bibinfo
  {author} {\bibfnamefont {J.~G.}\ \bibnamefont {Rarity}}, \bibinfo {author}
  {\bibfnamefont {S.}~\bibnamefont {Hughes}},\ and\ \bibinfo {author}
  {\bibfnamefont {R.}~\bibnamefont {Oulton}},\ }\bibfield  {title} {\bibinfo
  {title} {Polarization {{Engineering}} in {{Photonic Crystal Waveguides}} for
  {{Spin}}-{{Photon Entanglers}}},\ }\href
  {https://doi.org/10.1103/PhysRevLett.115.153901} {\bibfield  {journal}
  {\bibinfo  {journal} {Physical Review Letters}\ }\textbf {\bibinfo {volume}
  {115}},\ \bibinfo {pages} {153901} (\bibinfo {year} {2015})}\BibitemShut
  {NoStop}%
\bibitem [{\citenamefont {Mehrabad}\ \emph {et~al.}(2020)\citenamefont
  {Mehrabad}, \citenamefont {Foster}, \citenamefont {Dost}, \citenamefont
  {Clarke}, \citenamefont {Patil}, \citenamefont {Fox}, \citenamefont
  {Skolnick},\ and\ \citenamefont
  {Wilson}}]{mehrabadChiralTopologicalPhotonics2020}%
  \BibitemOpen
  \bibfield  {author} {\bibinfo {author} {\bibfnamefont {M.~J.}\ \bibnamefont
  {Mehrabad}}, \bibinfo {author} {\bibfnamefont {A.~P.}\ \bibnamefont
  {Foster}}, \bibinfo {author} {\bibfnamefont {R.}~\bibnamefont {Dost}},
  \bibinfo {author} {\bibfnamefont {E.}~\bibnamefont {Clarke}}, \bibinfo
  {author} {\bibfnamefont {P.~K.}\ \bibnamefont {Patil}}, \bibinfo {author}
  {\bibfnamefont {A.~M.}\ \bibnamefont {Fox}}, \bibinfo {author} {\bibfnamefont
  {M.~S.}\ \bibnamefont {Skolnick}},\ and\ \bibinfo {author} {\bibfnamefont
  {L.~R.}\ \bibnamefont {Wilson}},\ }\bibfield  {title} {\bibinfo {title}
  {Chiral topological photonics with an embedded quantum emitter},\ }\href
  {https://doi.org/10.1364/OPTICA.393035} {\bibfield  {journal} {\bibinfo
  {journal} {Optica}\ }\textbf {\bibinfo {volume} {7}},\ \bibinfo {pages}
  {1690} (\bibinfo {year} {2020})}\BibitemShut {NoStop}%
\bibitem [{\citenamefont {Su}\ \emph {et~al.}(2021)\citenamefont {Su},
  \citenamefont {Ghosh}, \citenamefont {Liew},\ and\ \citenamefont
  {Xiong}}]{Su2021}%
  \BibitemOpen
  \bibfield  {author} {\bibinfo {author} {\bibfnamefont {R.}~\bibnamefont
  {Su}}, \bibinfo {author} {\bibfnamefont {S.}~\bibnamefont {Ghosh}}, \bibinfo
  {author} {\bibfnamefont {T.~C.~H.}\ \bibnamefont {Liew}},\ and\ \bibinfo
  {author} {\bibfnamefont {Q.}~\bibnamefont {Xiong}},\ }\bibfield  {title}
  {\bibinfo {title} {Optical switching of topological phase in a perovskite
  polariton lattice},\ }\bibfield  {journal} {\bibinfo  {journal} {Science
  Advances}\ }\textbf {\bibinfo {volume} {7}},\ \href
  {https://doi.org/10.1126/sciadv.abf8049} {10.1126/sciadv.abf8049} (\bibinfo
  {year} {2021})\BibitemShut {NoStop}%
\bibitem [{\citenamefont {O'Faolain}\ \emph {et~al.}(2007)\citenamefont
  {O'Faolain}, \citenamefont {White}, \citenamefont {O'Brien}, \citenamefont
  {Yuan}, \citenamefont {Settle},\ and\ \citenamefont
  {Krauss}}]{ofaolainDependenceExtrinsicLoss2007}%
  \BibitemOpen
  \bibfield  {author} {\bibinfo {author} {\bibfnamefont {L.}~\bibnamefont
  {O'Faolain}}, \bibinfo {author} {\bibfnamefont {T.~P.}\ \bibnamefont
  {White}}, \bibinfo {author} {\bibfnamefont {D.}~\bibnamefont {O'Brien}},
  \bibinfo {author} {\bibfnamefont {X.}~\bibnamefont {Yuan}}, \bibinfo {author}
  {\bibfnamefont {M.~D.}\ \bibnamefont {Settle}},\ and\ \bibinfo {author}
  {\bibfnamefont {T.~F.}\ \bibnamefont {Krauss}},\ }\bibfield  {title}
  {\bibinfo {title} {Dependence of extrinsic loss on group velocity in photonic
  crystal waveguides},\ }\href {https://doi.org/10.1364/OE.15.013129}
  {\bibfield  {journal} {\bibinfo  {journal} {Optics Express}\ }\textbf
  {\bibinfo {volume} {15}},\ \bibinfo {pages} {13129} (\bibinfo {year}
  {2007})}\BibitemShut {NoStop}%
\bibitem [{\citenamefont {Hughes}\ \emph {et~al.}(2005)\citenamefont {Hughes},
  \citenamefont {Ramunno}, \citenamefont {Young},\ and\ \citenamefont
  {Sipe}}]{hughesExtrinsicOpticalScattering2005}%
  \BibitemOpen
  \bibfield  {author} {\bibinfo {author} {\bibfnamefont {S.}~\bibnamefont
  {Hughes}}, \bibinfo {author} {\bibfnamefont {L.}~\bibnamefont {Ramunno}},
  \bibinfo {author} {\bibfnamefont {J.~F.}\ \bibnamefont {Young}},\ and\
  \bibinfo {author} {\bibfnamefont {J.~E.}\ \bibnamefont {Sipe}},\ }\bibfield
  {title} {\bibinfo {title} {Extrinsic {{Optical Scattering Loss}} in
  {{Photonic Crystal Waveguides}}: {{Role}} of {{Fabrication Disorder}} and
  {{Photon Group Velocity}}},\ }\href
  {https://doi.org/10.1103/PhysRevLett.94.033903} {\bibfield  {journal}
  {\bibinfo  {journal} {Physical Review Letters}\ }\textbf {\bibinfo {volume}
  {94}},\ \bibinfo {pages} {033903} (\bibinfo {year} {2005})}\BibitemShut
  {NoStop}%
\bibitem [{\citenamefont {Patterson}\ \emph
  {et~al.}(2009{\natexlab{a}})\citenamefont {Patterson}, \citenamefont
  {Hughes}, \citenamefont {Combri{\'e}}, \citenamefont {Tran}, \citenamefont
  {De~Rossi}, \citenamefont {Gabet},\ and\ \citenamefont
  {Jaou{\"e}n}}]{pattersonDisorderInducedCoherentScattering2009}%
  \BibitemOpen
  \bibfield  {author} {\bibinfo {author} {\bibfnamefont {M.}~\bibnamefont
  {Patterson}}, \bibinfo {author} {\bibfnamefont {S.}~\bibnamefont {Hughes}},
  \bibinfo {author} {\bibfnamefont {S.}~\bibnamefont {Combri{\'e}}}, \bibinfo
  {author} {\bibfnamefont {N.-V.-Q.}\ \bibnamefont {Tran}}, \bibinfo {author}
  {\bibfnamefont {A.}~\bibnamefont {De~Rossi}}, \bibinfo {author}
  {\bibfnamefont {R.}~\bibnamefont {Gabet}},\ and\ \bibinfo {author}
  {\bibfnamefont {Y.}~\bibnamefont {Jaou{\"e}n}},\ }\bibfield  {title}
  {\bibinfo {title} {Disorder-{{Induced Coherent Scattering}} in
  {{Slow}}-{{Light Photonic Crystal Waveguides}}},\ }\href
  {https://doi.org/10.1103/PhysRevLett.102.253903} {\bibfield  {journal}
  {\bibinfo  {journal} {Physical Review Letters}\ }\textbf {\bibinfo {volume}
  {102}},\ \bibinfo {pages} {253903} (\bibinfo {year}
  {2009}{\natexlab{a}})}\BibitemShut {NoStop}%
\bibitem [{\citenamefont {Patterson}\ \emph
  {et~al.}(2009{\natexlab{b}})\citenamefont {Patterson}, \citenamefont
  {Hughes}, \citenamefont {Schulz}, \citenamefont {Beggs}, \citenamefont
  {White}, \citenamefont {O'Faolain},\ and\ \citenamefont
  {Krauss}}]{pattersonDisorderinducedIncoherentScattering2009}%
  \BibitemOpen
  \bibfield  {author} {\bibinfo {author} {\bibfnamefont {M.}~\bibnamefont
  {Patterson}}, \bibinfo {author} {\bibfnamefont {S.}~\bibnamefont {Hughes}},
  \bibinfo {author} {\bibfnamefont {S.}~\bibnamefont {Schulz}}, \bibinfo
  {author} {\bibfnamefont {D.~M.}\ \bibnamefont {Beggs}}, \bibinfo {author}
  {\bibfnamefont {T.~P.}\ \bibnamefont {White}}, \bibinfo {author}
  {\bibfnamefont {L.}~\bibnamefont {O'Faolain}},\ and\ \bibinfo {author}
  {\bibfnamefont {T.~F.}\ \bibnamefont {Krauss}},\ }\bibfield  {title}
  {\bibinfo {title} {Disorder-induced incoherent scattering losses in photonic
  crystal waveguides: {{Bloch}} mode reshaping, multiple scattering, and
  breakdown of the {{Beer}}-{{Lambert}} law},\ }\href
  {https://doi.org/10.1103/PhysRevB.80.195305} {\bibfield  {journal} {\bibinfo
  {journal} {Physical Review B}\ }\textbf {\bibinfo {volume} {80}},\ \bibinfo
  {pages} {195305} (\bibinfo {year} {2009}{\natexlab{b}})}\BibitemShut
  {NoStop}%
\bibitem [{\citenamefont {Andreani}\ and\ \citenamefont
  {Gerace}(2007)}]{andreaniLightMatterInteraction2007}%
  \BibitemOpen
  \bibfield  {author} {\bibinfo {author} {\bibfnamefont {L.~C.}\ \bibnamefont
  {Andreani}}\ and\ \bibinfo {author} {\bibfnamefont {D.}~\bibnamefont
  {Gerace}},\ }\bibfield  {title} {\bibinfo {title} {Light\textendash matter
  interaction in photonic crystal slabs},\ }\href
  {https://doi.org/10.1002/pssb.200743182} {\bibfield  {journal} {\bibinfo
  {journal} {physica status solidi (b)}\ }\textbf {\bibinfo {volume} {244}},\
  \bibinfo {pages} {3528} (\bibinfo {year} {2007})}\BibitemShut {NoStop}%
\bibitem [{\citenamefont {Kuramochi}\ \emph {et~al.}(2005)\citenamefont
  {Kuramochi}, \citenamefont {Notomi}, \citenamefont {Hughes}, \citenamefont
  {Shinya}, \citenamefont {Watanabe},\ and\ \citenamefont
  {Ramunno}}]{kuramochiDisorderinducedScatteringLoss2005}%
  \BibitemOpen
  \bibfield  {author} {\bibinfo {author} {\bibfnamefont {E.}~\bibnamefont
  {Kuramochi}}, \bibinfo {author} {\bibfnamefont {M.}~\bibnamefont {Notomi}},
  \bibinfo {author} {\bibfnamefont {S.}~\bibnamefont {Hughes}}, \bibinfo
  {author} {\bibfnamefont {A.}~\bibnamefont {Shinya}}, \bibinfo {author}
  {\bibfnamefont {T.}~\bibnamefont {Watanabe}},\ and\ \bibinfo {author}
  {\bibfnamefont {L.}~\bibnamefont {Ramunno}},\ }\bibfield  {title} {\bibinfo
  {title} {Disorder-induced scattering loss of line-defect waveguides in
  photonic crystal slabs},\ }\href {https://doi.org/10.1103/PhysRevB.72.161318}
  {\bibfield  {journal} {\bibinfo  {journal} {Physical Review B}\ }\textbf
  {\bibinfo {volume} {72}},\ \bibinfo {pages} {161318} (\bibinfo {year}
  {2005})}\BibitemShut {NoStop}%
\bibitem [{\citenamefont {Petrov}\ \emph {et~al.}(2009)\citenamefont {Petrov},
  \citenamefont {Krause},\ and\ \citenamefont
  {Eich}}]{petrovBackscatteringDisorderLimits2009}%
  \BibitemOpen
  \bibfield  {author} {\bibinfo {author} {\bibfnamefont {A.}~\bibnamefont
  {Petrov}}, \bibinfo {author} {\bibfnamefont {M.}~\bibnamefont {Krause}},\
  and\ \bibinfo {author} {\bibfnamefont {M.}~\bibnamefont {Eich}},\ }\bibfield
  {title} {\bibinfo {title} {Backscattering and disorder limits in slow light
  photonic crystal waveguides},\ }\href {https://doi.org/10.1364/OE.17.008676}
  {\bibfield  {journal} {\bibinfo  {journal} {Optics Express}\ }\textbf
  {\bibinfo {volume} {17}},\ \bibinfo {pages} {8676} (\bibinfo {year}
  {2009})}\BibitemShut {NoStop}%
\bibitem [{\citenamefont {Shalaev}\ \emph {et~al.}(2019)\citenamefont
  {Shalaev}, \citenamefont {Walasik}, \citenamefont {Tsukernik}, \citenamefont
  {Xu},\ and\ \citenamefont
  {Litchinitser}}]{shalaevRobustTopologicallyProtected2019}%
  \BibitemOpen
  \bibfield  {author} {\bibinfo {author} {\bibfnamefont {M.~I.}\ \bibnamefont
  {Shalaev}}, \bibinfo {author} {\bibfnamefont {W.}~\bibnamefont {Walasik}},
  \bibinfo {author} {\bibfnamefont {A.}~\bibnamefont {Tsukernik}}, \bibinfo
  {author} {\bibfnamefont {Y.}~\bibnamefont {Xu}},\ and\ \bibinfo {author}
  {\bibfnamefont {N.~M.}\ \bibnamefont {Litchinitser}},\ }\bibfield  {title}
  {\bibinfo {title} {Robust topologically protected transport in photonic
  crystals at telecommunication wavelengths},\ }\href
  {https://doi.org/10.1038/s41565-018-0297-6} {\bibfield  {journal} {\bibinfo
  {journal} {Nature Nanotechnology}\ }\textbf {\bibinfo {volume} {14}},\
  \bibinfo {pages} {31} (\bibinfo {year} {2019})}\BibitemShut {NoStop}%
\bibitem [{\citenamefont {He}\ \emph {et~al.}(2019)\citenamefont {He},
  \citenamefont {Liang}, \citenamefont {Yuan}, \citenamefont {Qiu},
  \citenamefont {Chen}, \citenamefont {Zhao},\ and\ \citenamefont
  {Dong}}]{heSilicononinsulatorSlabTopological2019}%
  \BibitemOpen
  \bibfield  {author} {\bibinfo {author} {\bibfnamefont {X.-T.}\ \bibnamefont
  {He}}, \bibinfo {author} {\bibfnamefont {E.-T.}\ \bibnamefont {Liang}},
  \bibinfo {author} {\bibfnamefont {J.-J.}\ \bibnamefont {Yuan}}, \bibinfo
  {author} {\bibfnamefont {H.-Y.}\ \bibnamefont {Qiu}}, \bibinfo {author}
  {\bibfnamefont {X.-D.}\ \bibnamefont {Chen}}, \bibinfo {author}
  {\bibfnamefont {F.-L.}\ \bibnamefont {Zhao}},\ and\ \bibinfo {author}
  {\bibfnamefont {J.-W.}\ \bibnamefont {Dong}},\ }\bibfield  {title} {\bibinfo
  {title} {A silicon-on-insulator slab for topological valley transport},\
  }\href {https://doi.org/10.1038/s41467-019-08881-z} {\bibfield  {journal}
  {\bibinfo  {journal} {Nature Communications}\ }\textbf {\bibinfo {volume}
  {10}},\ \bibinfo {pages} {872} (\bibinfo {year} {2019})}\BibitemShut
  {NoStop}%
\bibitem [{\citenamefont {Rosiek}\ \emph {et~al.}(2022)\citenamefont {Rosiek},
  \citenamefont {Arregui}, \citenamefont {Vladimirova}, \citenamefont
  {Albrechtsen}, \citenamefont {Lahijani}, \citenamefont {Christiansen},\ and\
  \citenamefont {Stobbe}}]{rosiekObservationStrongBackscattering2022}%
  \BibitemOpen
  \bibfield  {author} {\bibinfo {author} {\bibfnamefont {C.~A.}\ \bibnamefont
  {Rosiek}}, \bibinfo {author} {\bibfnamefont {G.}~\bibnamefont {Arregui}},
  \bibinfo {author} {\bibfnamefont {A.}~\bibnamefont {Vladimirova}}, \bibinfo
  {author} {\bibfnamefont {M.}~\bibnamefont {Albrechtsen}}, \bibinfo {author}
  {\bibfnamefont {B.~V.}\ \bibnamefont {Lahijani}}, \bibinfo {author}
  {\bibfnamefont {R.~E.}\ \bibnamefont {Christiansen}},\ and\ \bibinfo {author}
  {\bibfnamefont {S.}~\bibnamefont {Stobbe}},\ }\href@noop {} {\bibinfo {title}
  {Observation of strong backscattering in valley-{{Hall}} photonic topological
  interface modes}} (\bibinfo {year} {2022}),\ \Eprint
  {https://arxiv.org/abs/2206.11741} {arXiv:2206.11741 [physics]} \BibitemShut
  {NoStop}%
\bibitem [{\citenamefont {Hauff}\ \emph {et~al.}(2022)\citenamefont {Hauff},
  \citenamefont {Le~Jeannic}, \citenamefont {Lodahl}, \citenamefont {Hughes},\
  and\ \citenamefont {Rotenberg}}]{hauffChiralQuantumOptics}%
  \BibitemOpen
  \bibfield  {author} {\bibinfo {author} {\bibfnamefont {N.~V.}\ \bibnamefont
  {Hauff}}, \bibinfo {author} {\bibfnamefont {H.}~\bibnamefont {Le~Jeannic}},
  \bibinfo {author} {\bibfnamefont {P.}~\bibnamefont {Lodahl}}, \bibinfo
  {author} {\bibfnamefont {S.}~\bibnamefont {Hughes}},\ and\ \bibinfo {author}
  {\bibfnamefont {N.}~\bibnamefont {Rotenberg}},\ }\bibfield  {title} {\bibinfo
  {title} {Chiral quantum optics in broken-symmetry and topological photonic
  crystal waveguides},\ }\href
  {https://doi.org/10.1103/PhysRevResearch.4.023082} {\bibfield  {journal}
  {\bibinfo  {journal} {Phys. Rev. Research}\ }\textbf {\bibinfo {volume}
  {4}},\ \bibinfo {pages} {023082} (\bibinfo {year} {2022})}\BibitemShut
  {NoStop}%
\bibitem [{\citenamefont {Zhao}\ \emph {et~al.}(2015)\citenamefont {Zhao},
  \citenamefont {Zhang}, \citenamefont {Wang},\ and\ \citenamefont
  {Hu}}]{zhaoReviewOptimizationMethods2015}%
  \BibitemOpen
  \bibfield  {author} {\bibinfo {author} {\bibfnamefont {Y.}~\bibnamefont
  {Zhao}}, \bibinfo {author} {\bibfnamefont {Y.}~\bibnamefont {Zhang}},
  \bibinfo {author} {\bibfnamefont {Q.}~\bibnamefont {Wang}},\ and\ \bibinfo
  {author} {\bibfnamefont {H.}~\bibnamefont {Hu}},\ }\bibfield  {title}
  {\bibinfo {title} {Review on the {{Optimization Methods}} of {{Slow Light}}
  in {{Photonic Crystal Waveguide}}},\ }\href
  {https://doi.org/10.1109/TNANO.2015.2394410} {\bibfield  {journal} {\bibinfo
  {journal} {IEEE Transactions on Nanotechnology}\ }\textbf {\bibinfo {volume}
  {14}},\ \bibinfo {pages} {407} (\bibinfo {year} {2015})}\BibitemShut
  {NoStop}%
\bibitem [{\citenamefont {Li}\ \emph {et~al.}(2008)\citenamefont {Li},
  \citenamefont {White}, \citenamefont {O'Faolain}, \citenamefont
  {{Gomez-Iglesias}},\ and\ \citenamefont
  {Krauss}}]{liSystematicDesignFlat2008}%
  \BibitemOpen
  \bibfield  {author} {\bibinfo {author} {\bibfnamefont {J.}~\bibnamefont
  {Li}}, \bibinfo {author} {\bibfnamefont {T.~P.}\ \bibnamefont {White}},
  \bibinfo {author} {\bibfnamefont {L.}~\bibnamefont {O'Faolain}}, \bibinfo
  {author} {\bibfnamefont {A.}~\bibnamefont {{Gomez-Iglesias}}},\ and\ \bibinfo
  {author} {\bibfnamefont {T.~F.}\ \bibnamefont {Krauss}},\ }\bibfield  {title}
  {\bibinfo {title} {Systematic design of flat band slow light in photonic
  crystal waveguides},\ }\href {https://doi.org/10.1364/OE.16.006227}
  {\bibfield  {journal} {\bibinfo  {journal} {Optics Express}\ }\textbf
  {\bibinfo {volume} {16}},\ \bibinfo {pages} {6227} (\bibinfo {year}
  {2008})}\BibitemShut {NoStop}%
\bibitem [{\citenamefont {Li}\ \emph {et~al.}(2012)\citenamefont {Li},
  \citenamefont {O'Faolain}, \citenamefont {Schulz},\ and\ \citenamefont
  {Krauss}}]{liLowLossPropagation2012}%
  \BibitemOpen
  \bibfield  {author} {\bibinfo {author} {\bibfnamefont {J.}~\bibnamefont
  {Li}}, \bibinfo {author} {\bibfnamefont {L.}~\bibnamefont {O'Faolain}},
  \bibinfo {author} {\bibfnamefont {S.~A.}\ \bibnamefont {Schulz}},\ and\
  \bibinfo {author} {\bibfnamefont {T.~F.}\ \bibnamefont {Krauss}},\ }\bibfield
   {title} {\bibinfo {title} {Low loss propagation in slow light photonic
  crystal waveguides at group indices up to 60},\ }\href
  {https://doi.org/10.1016/j.photonics.2012.05.006} {\bibfield  {journal}
  {\bibinfo  {journal} {Photonics and Nanostructures - Fundamentals and
  Applications}\ }\bibinfo {series} {{{TaCoNa-Photonics}} 2011},\ \textbf
  {\bibinfo {volume} {10}},\ \bibinfo {pages} {589} (\bibinfo {year}
  {2012})}\BibitemShut {NoStop}%
\bibitem [{\citenamefont {Mori}\ and\ \citenamefont
  {Baba}(2005)}]{moriWidebandLowDispersion2005}%
  \BibitemOpen
  \bibfield  {author} {\bibinfo {author} {\bibfnamefont {D.}~\bibnamefont
  {Mori}}\ and\ \bibinfo {author} {\bibfnamefont {T.}~\bibnamefont {Baba}},\
  }\bibfield  {title} {\bibinfo {title} {Wideband and low dispersion slow light
  by chirped photonic crystal coupled waveguide},\ }\href
  {https://doi.org/10.1364/OPEX.13.009398} {\bibfield  {journal} {\bibinfo
  {journal} {Optics Express}\ }\textbf {\bibinfo {volume} {13}},\ \bibinfo
  {pages} {9398} (\bibinfo {year} {2005})}\BibitemShut {NoStop}%
\bibitem [{\citenamefont {Rigal}\ \emph {et~al.}(2017)\citenamefont {Rigal},
  \citenamefont {Joanesarson}, \citenamefont {Lyasota}, \citenamefont {Jarlov},
  \citenamefont {Dwir}, \citenamefont {Rudra}, \citenamefont {Kulkova},\ and\
  \citenamefont {Kapon}}]{rigalPropagationLossesPhotonic2017}%
  \BibitemOpen
  \bibfield  {author} {\bibinfo {author} {\bibfnamefont {B.}~\bibnamefont
  {Rigal}}, \bibinfo {author} {\bibfnamefont {K.}~\bibnamefont {Joanesarson}},
  \bibinfo {author} {\bibfnamefont {A.}~\bibnamefont {Lyasota}}, \bibinfo
  {author} {\bibfnamefont {C.}~\bibnamefont {Jarlov}}, \bibinfo {author}
  {\bibfnamefont {B.}~\bibnamefont {Dwir}}, \bibinfo {author} {\bibfnamefont
  {A.}~\bibnamefont {Rudra}}, \bibinfo {author} {\bibfnamefont
  {I.}~\bibnamefont {Kulkova}},\ and\ \bibinfo {author} {\bibfnamefont
  {E.}~\bibnamefont {Kapon}},\ }\bibfield  {title} {\bibinfo {title}
  {Propagation losses in photonic crystal waveguides: Effects of band tail
  absorption and waveguide dispersion},\ }\href
  {https://doi.org/10.1364/OE.25.028908} {\bibfield  {journal} {\bibinfo
  {journal} {Optics Express}\ }\textbf {\bibinfo {volume} {25}},\ \bibinfo
  {pages} {28908} (\bibinfo {year} {2017})}\BibitemShut {NoStop}%
\bibitem [{\citenamefont {Mahmoodian}\ \emph {et~al.}(2017)\citenamefont
  {Mahmoodian}, \citenamefont {{Prindal-Nielsen}}, \citenamefont {S{\"o}llner},
  \citenamefont {Stobbe},\ and\ \citenamefont
  {Lodahl}}]{mahmoodianEngineeringChiralLightmatter2017}%
  \BibitemOpen
  \bibfield  {author} {\bibinfo {author} {\bibfnamefont {S.}~\bibnamefont
  {Mahmoodian}}, \bibinfo {author} {\bibfnamefont {K.}~\bibnamefont
  {{Prindal-Nielsen}}}, \bibinfo {author} {\bibfnamefont {I.}~\bibnamefont
  {S{\"o}llner}}, \bibinfo {author} {\bibfnamefont {S.}~\bibnamefont
  {Stobbe}},\ and\ \bibinfo {author} {\bibfnamefont {P.}~\bibnamefont
  {Lodahl}},\ }\bibfield  {title} {\bibinfo {title} {Engineering chiral
  light-matter interaction in photonic crystal waveguides with slow light},\
  }\href {https://doi.org/10.1364/OME.7.000043} {\bibfield  {journal} {\bibinfo
   {journal} {Optical Materials Express}\ }\textbf {\bibinfo {volume} {7}},\
  \bibinfo {pages} {43} (\bibinfo {year} {2017})}\BibitemShut {NoStop}%
\bibitem [{\citenamefont {Molesky}\ \emph {et~al.}(2018)\citenamefont
  {Molesky}, \citenamefont {Lin}, \citenamefont {Piggott}, \citenamefont {Jin},
  \citenamefont {Vuckovi{\'c}},\ and\ \citenamefont
  {Rodriguez}}]{moleskyInverseDesignNanophotonics2018}%
  \BibitemOpen
  \bibfield  {author} {\bibinfo {author} {\bibfnamefont {S.}~\bibnamefont
  {Molesky}}, \bibinfo {author} {\bibfnamefont {Z.}~\bibnamefont {Lin}},
  \bibinfo {author} {\bibfnamefont {A.~Y.}\ \bibnamefont {Piggott}}, \bibinfo
  {author} {\bibfnamefont {W.}~\bibnamefont {Jin}}, \bibinfo {author}
  {\bibfnamefont {J.}~\bibnamefont {Vuckovi{\'c}}},\ and\ \bibinfo {author}
  {\bibfnamefont {A.~W.}\ \bibnamefont {Rodriguez}},\ }\bibfield  {title}
  {\bibinfo {title} {Inverse design in nanophotonics},\ }\href
  {https://doi.org/10.1038/s41566-018-0246-9} {\bibfield  {journal} {\bibinfo
  {journal} {Nature Photonics}\ }\textbf {\bibinfo {volume} {12}},\ \bibinfo
  {pages} {659} (\bibinfo {year} {2018})}\BibitemShut {NoStop}%
\bibitem [{\citenamefont {Minkov}\ \emph
  {et~al.}(2020{\natexlab{a}})\citenamefont {Minkov}, \citenamefont
  {Williamson}, \citenamefont {Andreani}, \citenamefont {Gerace}, \citenamefont
  {Lou}, \citenamefont {Song}, \citenamefont {Hughes},\ and\ \citenamefont
  {Fan}}]{minkovInverseDesignPhotonic2020}%
  \BibitemOpen
  \bibfield  {author} {\bibinfo {author} {\bibfnamefont {M.}~\bibnamefont
  {Minkov}}, \bibinfo {author} {\bibfnamefont {I.~A.~D.}\ \bibnamefont
  {Williamson}}, \bibinfo {author} {\bibfnamefont {L.~C.}\ \bibnamefont
  {Andreani}}, \bibinfo {author} {\bibfnamefont {D.}~\bibnamefont {Gerace}},
  \bibinfo {author} {\bibfnamefont {B.}~\bibnamefont {Lou}}, \bibinfo {author}
  {\bibfnamefont {A.~Y.}\ \bibnamefont {Song}}, \bibinfo {author}
  {\bibfnamefont {T.~W.}\ \bibnamefont {Hughes}},\ and\ \bibinfo {author}
  {\bibfnamefont {S.}~\bibnamefont {Fan}},\ }\bibfield  {title} {\bibinfo
  {title} {Inverse {{Design}} of {{Photonic Crystals}} through {{Automatic
  Differentiation}}},\ }\href {https://doi.org/10.1021/acsphotonics.0c00327}
  {\bibfield  {journal} {\bibinfo  {journal} {ACS Photonics}\ }\textbf
  {\bibinfo {volume} {7}},\ \bibinfo {pages} {1729} (\bibinfo {year}
  {2020}{\natexlab{a}})}\BibitemShut {NoStop}%
\bibitem [{\citenamefont {Piggott}\ \emph {et~al.}(2015)\citenamefont
  {Piggott}, \citenamefont {Lu}, \citenamefont {Lagoudakis}, \citenamefont
  {Petykiewicz}, \citenamefont {Babinec},\ and\ \citenamefont {Vu{\v
  c}kovi{\'c}}}]{piggottInverseDesignDemonstration2015}%
  \BibitemOpen
  \bibfield  {author} {\bibinfo {author} {\bibfnamefont {A.~Y.}\ \bibnamefont
  {Piggott}}, \bibinfo {author} {\bibfnamefont {J.}~\bibnamefont {Lu}},
  \bibinfo {author} {\bibfnamefont {K.~G.}\ \bibnamefont {Lagoudakis}},
  \bibinfo {author} {\bibfnamefont {J.}~\bibnamefont {Petykiewicz}}, \bibinfo
  {author} {\bibfnamefont {T.~M.}\ \bibnamefont {Babinec}},\ and\ \bibinfo
  {author} {\bibfnamefont {J.}~\bibnamefont {Vu{\v c}kovi{\'c}}},\ }\bibfield
  {title} {\bibinfo {title} {Inverse design and demonstration of a compact and
  broadband on-chip wavelength demultiplexer},\ }\href
  {https://doi.org/10.1038/nphoton.2015.69} {\bibfield  {journal} {\bibinfo
  {journal} {Nature Photonics}\ }\textbf {\bibinfo {volume} {9}},\ \bibinfo
  {pages} {374} (\bibinfo {year} {2015})}\BibitemShut {NoStop}%
\bibitem [{\citenamefont {Christiansen}\ \emph {et~al.}(2019)\citenamefont
  {Christiansen}, \citenamefont {Wang},\ and\ \citenamefont
  {Sigmund}}]{christiansenTopologicalInsulatorsTopology2019}%
  \BibitemOpen
  \bibfield  {author} {\bibinfo {author} {\bibfnamefont {R.~E.}\ \bibnamefont
  {Christiansen}}, \bibinfo {author} {\bibfnamefont {F.}~\bibnamefont {Wang}},\
  and\ \bibinfo {author} {\bibfnamefont {O.}~\bibnamefont {Sigmund}},\
  }\bibfield  {title} {\bibinfo {title} {Topological {{Insulators}} by
  {{Topology Optimization}}},\ }\href
  {https://doi.org/10.1103/PhysRevLett.122.234502} {\bibfield  {journal}
  {\bibinfo  {journal} {Physical Review Letters}\ }\textbf {\bibinfo {volume}
  {122}},\ \bibinfo {pages} {234502} (\bibinfo {year} {2019})}\BibitemShut
  {NoStop}%
\bibitem [{\citenamefont {Mann}\ \emph {et~al.}(2013)\citenamefont {Mann},
  \citenamefont {Combri{\'e}}, \citenamefont {Colman}, \citenamefont
  {Patterson}, \citenamefont {Rossi},\ and\ \citenamefont
  {Hughes}}]{mannReducingDisorderinducedLosses2013}%
  \BibitemOpen
  \bibfield  {author} {\bibinfo {author} {\bibfnamefont {N.}~\bibnamefont
  {Mann}}, \bibinfo {author} {\bibfnamefont {S.}~\bibnamefont {Combri{\'e}}},
  \bibinfo {author} {\bibfnamefont {P.}~\bibnamefont {Colman}}, \bibinfo
  {author} {\bibfnamefont {M.}~\bibnamefont {Patterson}}, \bibinfo {author}
  {\bibfnamefont {A.~D.}\ \bibnamefont {Rossi}},\ and\ \bibinfo {author}
  {\bibfnamefont {S.}~\bibnamefont {Hughes}},\ }\bibfield  {title} {\bibinfo
  {title} {Reducing disorder-induced losses for slow light photonic crystal
  waveguides through {{Bloch}} mode engineering},\ }\href
  {https://doi.org/10.1364/OL.38.004244} {\bibfield  {journal} {\bibinfo
  {journal} {Optics Letters}\ }\textbf {\bibinfo {volume} {38}},\ \bibinfo
  {pages} {4244} (\bibinfo {year} {2013})}\BibitemShut {NoStop}%
\bibitem [{\citenamefont {Andreani}\ and\ \citenamefont
  {Gerace}(2006)}]{andreaniPhotoniccrystalSlabsTriangular2006}%
  \BibitemOpen
  \bibfield  {author} {\bibinfo {author} {\bibfnamefont {L.~C.}\ \bibnamefont
  {Andreani}}\ and\ \bibinfo {author} {\bibfnamefont {D.}~\bibnamefont
  {Gerace}},\ }\bibfield  {title} {\bibinfo {title} {Photonic-crystal slabs
  with a triangular lattice of triangular holes investigated using a
  guided-mode expansion method},\ }\href
  {https://doi.org/10.1103/PhysRevB.73.235114} {\bibfield  {journal} {\bibinfo
  {journal} {Physical Review B}\ }\textbf {\bibinfo {volume} {73}},\ \bibinfo
  {pages} {235114} (\bibinfo {year} {2006})}\BibitemShut {NoStop}%
\bibitem [{\citenamefont {Sauer}(2021)}]{sauerTheoryComputationIntrinsic2021}%
  \BibitemOpen
  \bibfield  {author} {\bibinfo {author} {\bibfnamefont {E.}~\bibnamefont
  {Sauer}},\ }\emph {\bibinfo {title} {Theory and {{Computation}} of
  {{Intrinsic Propagation Losses}} and {{Disorder}}-{{Induced Modes}} in
  {{Topological Photonic Crystal Waveguides Using Mode Expansion
  Techniques}}}},\ \href@noop {} {Ph.D. thesis},\ \bibinfo  {school} {Queen's
  University}, \bibinfo {address} {{Kingston, Ontario, Canada}} (\bibinfo
  {year} {2021})\BibitemShut {NoStop}%
\bibitem [{\citenamefont {Vasco}\ and\ \citenamefont
  {Hughes}(2017)}]{vascoStatisticsAndersonlocalizedModes2017}%
  \BibitemOpen
  \bibfield  {author} {\bibinfo {author} {\bibfnamefont {J.~P.}\ \bibnamefont
  {Vasco}}\ and\ \bibinfo {author} {\bibfnamefont {S.}~\bibnamefont {Hughes}},\
  }\bibfield  {title} {\bibinfo {title} {Statistics of {{Anderson}}-localized
  modes in disordered photonic crystal slab waveguides},\ }\href
  {https://doi.org/10.1103/PhysRevB.95.224202} {\bibfield  {journal} {\bibinfo
  {journal} {Physical Review B}\ }\textbf {\bibinfo {volume} {95}},\ \bibinfo
  {pages} {224202} (\bibinfo {year} {2017})}\BibitemShut {NoStop}%
\bibitem [{\citenamefont {Sauer}\ \emph {et~al.}(2020)\citenamefont {Sauer},
  \citenamefont {Vasco},\ and\ \citenamefont
  {Hughes}}]{sauerTheoryIntrinsicPropagation2020}%
  \BibitemOpen
  \bibfield  {author} {\bibinfo {author} {\bibfnamefont {E.}~\bibnamefont
  {Sauer}}, \bibinfo {author} {\bibfnamefont {J.~P.}\ \bibnamefont {Vasco}},\
  and\ \bibinfo {author} {\bibfnamefont {S.}~\bibnamefont {Hughes}},\
  }\bibfield  {title} {\bibinfo {title} {Theory of intrinsic propagation losses
  in topological edge states of planar photonic crystals},\ }\href
  {https://doi.org/10.1103/PhysRevResearch.2.043109} {\bibfield  {journal}
  {\bibinfo  {journal} {Physical Review Research}\ }\textbf {\bibinfo {volume}
  {2}},\ \bibinfo {pages} {043109} (\bibinfo {year} {2020})}\BibitemShut
  {NoStop}%
\bibitem [{\citenamefont {Minkov}\ \emph
  {et~al.}(2020{\natexlab{b}})\citenamefont {Minkov}, \citenamefont
  {Williamson},\ and\ \citenamefont {Fan}}]{FancomputeLegume2020}%
  \BibitemOpen
  \bibfield  {author} {\bibinfo {author} {\bibfnamefont {M.}~\bibnamefont
  {Minkov}}, \bibinfo {author} {\bibfnamefont {I.}~\bibnamefont {Williamson}},\
  and\ \bibinfo {author} {\bibfnamefont {S.}~\bibnamefont {Fan}},\ }\href
  {https://legume.readthedocs.io/en/latest/index.html} {\bibinfo {title}
  {legume: Differentiable guided mode expansion methods}},\ \bibinfo
  {howpublished} {GitHub, \url{https://github.com/fancompute/legume}} (\bibinfo
  {year} {2020}{\natexlab{b}})\BibitemShut {NoStop}%
\bibitem [{\citenamefont
  {Andreani}(2006)}]{andreaniGuidedModeExpansionMethod2006}%
  \BibitemOpen
  \bibfield  {author} {\bibinfo {author} {\bibfnamefont {L.~C.}\ \bibnamefont
  {Andreani}},\ }\href@noop {} {\bibinfo {title} {The {{Guided}}-{{Mode
  Expansion Method}} for {{Photonic Crystal Slabs}}}} (\bibinfo {year}
  {2006})\BibitemShut {NoStop}%
\bibitem [{\citenamefont {Franke}\ \emph {et~al.}(2021)\citenamefont {Franke},
  \citenamefont {Ren}, \citenamefont {Richter}, \citenamefont {Knorr},\ and\
  \citenamefont {Hughes}}]{PhysRevLett.127.013602}%
  \BibitemOpen
  \bibfield  {author} {\bibinfo {author} {\bibfnamefont {S.}~\bibnamefont
  {Franke}}, \bibinfo {author} {\bibfnamefont {J.}~\bibnamefont {Ren}},
  \bibinfo {author} {\bibfnamefont {M.}~\bibnamefont {Richter}}, \bibinfo
  {author} {\bibfnamefont {A.}~\bibnamefont {Knorr}},\ and\ \bibinfo {author}
  {\bibfnamefont {S.}~\bibnamefont {Hughes}},\ }\bibfield  {title} {\bibinfo
  {title} {Fermi's golden rule for spontaneous emission in absorptive and
  amplifying media},\ }\href {https://doi.org/10.1103/PhysRevLett.127.013602}
  {\bibfield  {journal} {\bibinfo  {journal} {Phys. Rev. Lett.}\ }\textbf
  {\bibinfo {volume} {127}},\ \bibinfo {pages} {013602} (\bibinfo {year}
  {2021})}\BibitemShut {NoStop}%
\bibitem [{\citenamefont {Maclaurin}\ \emph {et~al.}()\citenamefont
  {Maclaurin}, \citenamefont {Duvenaud}, \citenamefont {Johnson},\ and\
  \citenamefont {Townsend}}]{maclaurinHIPSAutograd}%
  \BibitemOpen
  \bibfield  {author} {\bibinfo {author} {\bibfnamefont {D.}~\bibnamefont
  {Maclaurin}}, \bibinfo {author} {\bibfnamefont {D.}~\bibnamefont {Duvenaud}},
  \bibinfo {author} {\bibfnamefont {M.}~\bibnamefont {Johnson}},\ and\ \bibinfo
  {author} {\bibfnamefont {J.}~\bibnamefont {Townsend}},\ }\href
  {https://github.com/HIPS/autograd} {\bibinfo {title} {Autograd}},\ \bibinfo
  {howpublished} {GitHub, \url{https://github.com/HIPS/autograd}}\BibitemShut
  {NoStop}%
\bibitem [{\citenamefont {Michaels}\ and\ \citenamefont
  {Yablonovitch}(2018)}]{michaelsLeveragingContinuousMaterial2018}%
  \BibitemOpen
  \bibfield  {author} {\bibinfo {author} {\bibfnamefont {A.}~\bibnamefont
  {Michaels}}\ and\ \bibinfo {author} {\bibfnamefont {E.}~\bibnamefont
  {Yablonovitch}},\ }\bibfield  {title} {\bibinfo {title} {Leveraging
  continuous material averaging for inverse electromagnetic design},\ }\href
  {https://doi.org/10.1364/OE.26.031717} {\bibfield  {journal} {\bibinfo
  {journal} {Optics Express}\ }\textbf {\bibinfo {volume} {26}},\ \bibinfo
  {pages} {31717} (\bibinfo {year} {2018})}\BibitemShut {NoStop}%
\bibitem [{\citenamefont {Yoshimi}\ \emph {et~al.}(2020)\citenamefont
  {Yoshimi}, \citenamefont {Yamaguchi}, \citenamefont {Ota}, \citenamefont
  {Arakawa},\ and\ \citenamefont {Iwamoto}}]{yoshimiSlowLightWaveguides2020}%
  \BibitemOpen
  \bibfield  {author} {\bibinfo {author} {\bibfnamefont {H.}~\bibnamefont
  {Yoshimi}}, \bibinfo {author} {\bibfnamefont {T.}~\bibnamefont {Yamaguchi}},
  \bibinfo {author} {\bibfnamefont {Y.}~\bibnamefont {Ota}}, \bibinfo {author}
  {\bibfnamefont {Y.}~\bibnamefont {Arakawa}},\ and\ \bibinfo {author}
  {\bibfnamefont {S.}~\bibnamefont {Iwamoto}},\ }\bibfield  {title} {\bibinfo
  {title} {Slow light waveguides in topological valley photonic crystals},\
  }\href {https://doi.org/10.1364/OL.391764} {\bibfield  {journal} {\bibinfo
  {journal} {Optics Letters}\ }\textbf {\bibinfo {volume} {45}},\ \bibinfo
  {pages} {2648} (\bibinfo {year} {2020})}\BibitemShut {NoStop}%
\bibitem [{\citenamefont {Yoshimi}\ \emph {et~al.}(2021)\citenamefont
  {Yoshimi}, \citenamefont {Yamaguchi}, \citenamefont {Katsumi}, \citenamefont
  {Ota}, \citenamefont {Arakawa},\ and\ \citenamefont
  {Iwamoto}}]{yoshimiExperimentalDemonstrationTopological2021}%
  \BibitemOpen
  \bibfield  {author} {\bibinfo {author} {\bibfnamefont {H.}~\bibnamefont
  {Yoshimi}}, \bibinfo {author} {\bibfnamefont {T.}~\bibnamefont {Yamaguchi}},
  \bibinfo {author} {\bibfnamefont {R.}~\bibnamefont {Katsumi}}, \bibinfo
  {author} {\bibfnamefont {Y.}~\bibnamefont {Ota}}, \bibinfo {author}
  {\bibfnamefont {Y.}~\bibnamefont {Arakawa}},\ and\ \bibinfo {author}
  {\bibfnamefont {S.}~\bibnamefont {Iwamoto}},\ }\bibfield  {title} {\bibinfo
  {title} {Experimental demonstration of topological slow light waveguides in
  valley photonic crystals},\ }\href {https://doi.org/10.1364/OE.422962}
  {\bibfield  {journal} {\bibinfo  {journal} {Optics Express}\ }\textbf
  {\bibinfo {volume} {29}},\ \bibinfo {pages} {13441} (\bibinfo {year}
  {2021})}\BibitemShut {NoStop}%
\bibitem [{\citenamefont {Barik}\ \emph {et~al.}(2018)\citenamefont {Barik},
  \citenamefont {Karasahin}, \citenamefont {Flower}, \citenamefont {Cai},
  \citenamefont {Miyake}, \citenamefont {DeGottardi}, \citenamefont {Hafezi},\
  and\ \citenamefont {Waks}}]{barikTopologicalQuantumOptics2018}%
  \BibitemOpen
  \bibfield  {author} {\bibinfo {author} {\bibfnamefont {S.}~\bibnamefont
  {Barik}}, \bibinfo {author} {\bibfnamefont {A.}~\bibnamefont {Karasahin}},
  \bibinfo {author} {\bibfnamefont {C.}~\bibnamefont {Flower}}, \bibinfo
  {author} {\bibfnamefont {T.}~\bibnamefont {Cai}}, \bibinfo {author}
  {\bibfnamefont {H.}~\bibnamefont {Miyake}}, \bibinfo {author} {\bibfnamefont
  {W.}~\bibnamefont {DeGottardi}}, \bibinfo {author} {\bibfnamefont
  {M.}~\bibnamefont {Hafezi}},\ and\ \bibinfo {author} {\bibfnamefont
  {E.}~\bibnamefont {Waks}},\ }\bibfield  {title} {\bibinfo {title} {A
  topological quantum optics interface},\ }\href
  {https://doi.org/10.1126/science.aaq0327} {\bibfield  {journal} {\bibinfo
  {journal} {Science}\ }\textbf {\bibinfo {volume} {359}},\ \bibinfo {pages}
  {666} (\bibinfo {year} {2018})}\BibitemShut {NoStop}%
\bibitem [{\citenamefont {Lu}\ \emph {et~al.}(2014)\citenamefont {Lu},
  \citenamefont {Joannopoulos},\ and\ \citenamefont {Solja{\v
  c}i{\'c}}}]{luTopologicalPhotonics2014}%
  \BibitemOpen
  \bibfield  {author} {\bibinfo {author} {\bibfnamefont {L.}~\bibnamefont
  {Lu}}, \bibinfo {author} {\bibfnamefont {J.~D.}\ \bibnamefont
  {Joannopoulos}},\ and\ \bibinfo {author} {\bibfnamefont {M.}~\bibnamefont
  {Solja{\v c}i{\'c}}},\ }\bibfield  {title} {\bibinfo {title} {Topological
  photonics},\ }\href {https://doi.org/10.1038/nphoton.2014.248} {\bibfield
  {journal} {\bibinfo  {journal} {Nature Photonics}\ }\textbf {\bibinfo
  {volume} {8}},\ \bibinfo {pages} {821} (\bibinfo {year} {2014})}\BibitemShut
  {NoStop}%
\bibitem [{\citenamefont {de~Paz}\ \emph {et~al.}(2020)\citenamefont {de~Paz},
  \citenamefont {Devescovi}, \citenamefont {Giedke}, \citenamefont {Saenz},
  \citenamefont {Vergniory}, \citenamefont {Bradlyn}, \citenamefont
  {Bercioux},\ and\ \citenamefont
  {{Garc{\'i}a-Etxarri}}}]{pazTutorialComputingTopological2020}%
  \BibitemOpen
  \bibfield  {author} {\bibinfo {author} {\bibfnamefont {M.~B.}\ \bibnamefont
  {de~Paz}}, \bibinfo {author} {\bibfnamefont {C.}~\bibnamefont {Devescovi}},
  \bibinfo {author} {\bibfnamefont {G.}~\bibnamefont {Giedke}}, \bibinfo
  {author} {\bibfnamefont {J.~J.}\ \bibnamefont {Saenz}}, \bibinfo {author}
  {\bibfnamefont {M.~G.}\ \bibnamefont {Vergniory}}, \bibinfo {author}
  {\bibfnamefont {B.}~\bibnamefont {Bradlyn}}, \bibinfo {author} {\bibfnamefont
  {D.}~\bibnamefont {Bercioux}},\ and\ \bibinfo {author} {\bibfnamefont
  {A.}~\bibnamefont {{Garc{\'i}a-Etxarri}}},\ }\bibfield  {title} {\bibinfo
  {title} {Tutorial: {{Computing Topological Invariants}} in {{2D Photonic
  Crystals}}},\ }\href {https://doi.org/10.1002/qute.201900117} {\bibfield
  {journal} {\bibinfo  {journal} {Advanced Quantum Technologies}\ }\textbf
  {\bibinfo {volume} {3}},\ \bibinfo {pages} {1900117} (\bibinfo {year}
  {2020})}\BibitemShut {NoStop}%
\bibitem [{\citenamefont {Lodahl}\ \emph
  {et~al.}(2015{\natexlab{b}})\citenamefont {Lodahl}, \citenamefont
  {Mahmoodian},\ and\ \citenamefont
  {Stobbe}}]{lodahlInterfacingSinglePhotons2015}%
  \BibitemOpen
  \bibfield  {author} {\bibinfo {author} {\bibfnamefont {P.}~\bibnamefont
  {Lodahl}}, \bibinfo {author} {\bibfnamefont {S.}~\bibnamefont {Mahmoodian}},\
  and\ \bibinfo {author} {\bibfnamefont {S.}~\bibnamefont {Stobbe}},\
  }\bibfield  {title} {\bibinfo {title} {Interfacing single photons and single
  quantum dots with photonic nanostructures},\ }\href
  {https://doi.org/10.1103/RevModPhys.87.347} {\bibfield  {journal} {\bibinfo
  {journal} {Reviews of Modern Physics}\ }\textbf {\bibinfo {volume} {87}},\
  \bibinfo {pages} {347} (\bibinfo {year} {2015}{\natexlab{b}})}\BibitemShut
  {NoStop}%
\bibitem [{\citenamefont {Pregnolato}\ \emph {et~al.}(2020)\citenamefont
  {Pregnolato}, \citenamefont {Chu}, \citenamefont {Schr{\"o}der},
  \citenamefont {Schott}, \citenamefont {Wieck}, \citenamefont {Ludwig},
  \citenamefont {Lodahl},\ and\ \citenamefont
  {Rotenberg}}]{pregnolatoDeterministicPositioningNanophotonic2020}%
  \BibitemOpen
  \bibfield  {author} {\bibinfo {author} {\bibfnamefont {T.}~\bibnamefont
  {Pregnolato}}, \bibinfo {author} {\bibfnamefont {X.-L.}\ \bibnamefont {Chu}},
  \bibinfo {author} {\bibfnamefont {T.}~\bibnamefont {Schr{\"o}der}}, \bibinfo
  {author} {\bibfnamefont {R.}~\bibnamefont {Schott}}, \bibinfo {author}
  {\bibfnamefont {A.~D.}\ \bibnamefont {Wieck}}, \bibinfo {author}
  {\bibfnamefont {A.}~\bibnamefont {Ludwig}}, \bibinfo {author} {\bibfnamefont
  {P.}~\bibnamefont {Lodahl}},\ and\ \bibinfo {author} {\bibfnamefont
  {N.}~\bibnamefont {Rotenberg}},\ }\bibfield  {title} {\bibinfo {title}
  {Deterministic positioning of nanophotonic waveguides around single
  self-assembled quantum dots},\ }\href {https://doi.org/10.1063/1.5117888}
  {\bibfield  {journal} {\bibinfo  {journal} {APL Photonics}\ }\textbf
  {\bibinfo {volume} {5}},\ \bibinfo {pages} {086101} (\bibinfo {year}
  {2020})}\BibitemShut {NoStop}%
\bibitem [{\citenamefont {Gustin}\ and\ \citenamefont
  {Hughes}(2018)}]{PhysRevB.98.045309}%
  \BibitemOpen
  \bibfield  {author} {\bibinfo {author} {\bibfnamefont {C.}~\bibnamefont
  {Gustin}}\ and\ \bibinfo {author} {\bibfnamefont {S.}~\bibnamefont
  {Hughes}},\ }\bibfield  {title} {\bibinfo {title} {Pulsed excitation dynamics
  in quantum-dot--cavity systems: Limits to optimizing the fidelity of
  on-demand single-photon sources},\ }\href
  {https://doi.org/10.1103/PhysRevB.98.045309} {\bibfield  {journal} {\bibinfo
  {journal} {Phys. Rev. B}\ }\textbf {\bibinfo {volume} {98}},\ \bibinfo
  {pages} {045309} (\bibinfo {year} {2018})}\BibitemShut {NoStop}%
\bibitem [{\citenamefont {Somaschi}\ \emph {et~al.}(2016)\citenamefont
  {Somaschi}, \citenamefont {Giesz}, \citenamefont {Santis}, \citenamefont
  {Loredo}, \citenamefont {Almeida}, \citenamefont {Hornecker}, \citenamefont
  {Portalupi}, \citenamefont {Grange}, \citenamefont {Ant{\'{o}}n},
  \citenamefont {Demory}, \citenamefont {G{\'{o}}mez}, \citenamefont {Sagnes},
  \citenamefont {Lanzillotti-Kimura}, \citenamefont {Lema{\'{\i}}tre},
  \citenamefont {Auffeves}, \citenamefont {White}, \citenamefont {Lanco},\ and\
  \citenamefont {Senellart}}]{Somaschi2016}%
  \BibitemOpen
  \bibfield  {author} {\bibinfo {author} {\bibfnamefont {N.}~\bibnamefont
  {Somaschi}}, \bibinfo {author} {\bibfnamefont {V.}~\bibnamefont {Giesz}},
  \bibinfo {author} {\bibfnamefont {L.~D.}\ \bibnamefont {Santis}}, \bibinfo
  {author} {\bibfnamefont {J.~C.}\ \bibnamefont {Loredo}}, \bibinfo {author}
  {\bibfnamefont {M.~P.}\ \bibnamefont {Almeida}}, \bibinfo {author}
  {\bibfnamefont {G.}~\bibnamefont {Hornecker}}, \bibinfo {author}
  {\bibfnamefont {S.~L.}\ \bibnamefont {Portalupi}}, \bibinfo {author}
  {\bibfnamefont {T.}~\bibnamefont {Grange}}, \bibinfo {author} {\bibfnamefont
  {C.}~\bibnamefont {Ant{\'{o}}n}}, \bibinfo {author} {\bibfnamefont
  {J.}~\bibnamefont {Demory}}, \bibinfo {author} {\bibfnamefont
  {C.}~\bibnamefont {G{\'{o}}mez}}, \bibinfo {author} {\bibfnamefont
  {I.}~\bibnamefont {Sagnes}}, \bibinfo {author} {\bibfnamefont {N.~D.}\
  \bibnamefont {Lanzillotti-Kimura}}, \bibinfo {author} {\bibfnamefont
  {A.}~\bibnamefont {Lema{\'{\i}}tre}}, \bibinfo {author} {\bibfnamefont
  {A.}~\bibnamefont {Auffeves}}, \bibinfo {author} {\bibfnamefont {A.~G.}\
  \bibnamefont {White}}, \bibinfo {author} {\bibfnamefont {L.}~\bibnamefont
  {Lanco}},\ and\ \bibinfo {author} {\bibfnamefont {P.}~\bibnamefont
  {Senellart}},\ }\bibfield  {title} {\bibinfo {title} {Near-optimal
  single-photon sources in the solid state},\ }\href
  {https://doi.org/10.1038/nphoton.2016.23} {\bibfield  {journal} {\bibinfo
  {journal} {Nature Photonics}\ }\textbf {\bibinfo {volume} {10}},\ \bibinfo
  {pages} {340} (\bibinfo {year} {2016})}\BibitemShut {NoStop}%
\bibitem [{\citenamefont {Ding}\ \emph {et~al.}(2016)\citenamefont {Ding},
  \citenamefont {He}, \citenamefont {Duan}, \citenamefont {Gregersen},
  \citenamefont {Chen}, \citenamefont {Unsleber}, \citenamefont {Maier},
  \citenamefont {Schneider}, \citenamefont {Kamp}, \citenamefont {H\"ofling},
  \citenamefont {Lu},\ and\ \citenamefont {Pan}}]{PhysRevLett.116.020401}%
  \BibitemOpen
  \bibfield  {author} {\bibinfo {author} {\bibfnamefont {X.}~\bibnamefont
  {Ding}}, \bibinfo {author} {\bibfnamefont {Y.}~\bibnamefont {He}}, \bibinfo
  {author} {\bibfnamefont {Z.-C.}\ \bibnamefont {Duan}}, \bibinfo {author}
  {\bibfnamefont {N.}~\bibnamefont {Gregersen}}, \bibinfo {author}
  {\bibfnamefont {M.-C.}\ \bibnamefont {Chen}}, \bibinfo {author}
  {\bibfnamefont {S.}~\bibnamefont {Unsleber}}, \bibinfo {author}
  {\bibfnamefont {S.}~\bibnamefont {Maier}}, \bibinfo {author} {\bibfnamefont
  {C.}~\bibnamefont {Schneider}}, \bibinfo {author} {\bibfnamefont
  {M.}~\bibnamefont {Kamp}}, \bibinfo {author} {\bibfnamefont {S.}~\bibnamefont
  {H\"ofling}}, \bibinfo {author} {\bibfnamefont {C.-Y.}\ \bibnamefont {Lu}},\
  and\ \bibinfo {author} {\bibfnamefont {J.-W.}\ \bibnamefont {Pan}},\
  }\bibfield  {title} {\bibinfo {title} {On-demand single photons with high
  extraction efficiency and near-unity indistinguishability from a resonantly
  driven quantum dot in a micropillar},\ }\href
  {https://doi.org/10.1103/PhysRevLett.116.020401} {\bibfield  {journal}
  {\bibinfo  {journal} {Phys. Rev. Lett.}\ }\textbf {\bibinfo {volume} {116}},\
  \bibinfo {pages} {020401} (\bibinfo {year} {2016})}\BibitemShut {NoStop}%
\bibitem [{\citenamefont {Kingma}\ and\ \citenamefont
  {Ba}(2017)}]{kingmaAdamMethodStochastic2017}%
  \BibitemOpen
  \bibfield  {author} {\bibinfo {author} {\bibfnamefont {D.~P.}\ \bibnamefont
  {Kingma}}\ and\ \bibinfo {author} {\bibfnamefont {J.}~\bibnamefont {Ba}},\
  }\bibfield  {title} {\bibinfo {title} {Adam: {{A Method}} for {{Stochastic
  Optimization}}},\ }in\ \href@noop {} {\emph {\bibinfo {booktitle}
  {{{arXiv}}:1412.6980 [Cs]}}}\ (\bibinfo {year} {2017})\ \Eprint
  {https://arxiv.org/abs/1412.6980} {arXiv:1412.6980 [cs]} \BibitemShut
  {NoStop}%
\end{thebibliography}%

\end{document}